\documentclass[sigconf]{acmart} 
\newcommand{\revision}[1]{{ #1}}
\AtBeginDocument{%
  \providecommand\BibTeX{{%
    \normalfont B\kern-0.5em{\scshape i\kern-0.25em b}\kern-0.8em\TeX}}}

\setcopyright{none}
\renewcommand\footnotetextcopyrightpermission[1]{} 
\begin{document}

\title[Identifying the Gap in Empowering AI to Participate Equally in Group Decision-Making]{Competent but Rigid: Identifying the Gap in Empowering AI to Participate Equally in Group Decision-Making}

\author{Chengbo Zheng}
\email{cb.zheng@connect.ust.hk}
\orcid{0000-0003-0226-9399}
\affiliation{%
  \institution{The Hong Kong University of Science and Technology}
  \city{Hong Kong}
  \country{China}
}

\author{Yuheng Wu}
\email{220041080@link.cuhk.edu.cn}
\orcid{0000-0001-8473-0999}
\affiliation{%
  \institution{Chinese University of Hong Kong, Shenzhen}
  \city{Shenzhen}
  \country{China}
}

\author{Chuhan Shi}
\email{cshiag@connect.ust.hk}
\orcid{0000-0002-3370-1626}
\affiliation{%
  \institution{The Hong Kong University of Science and Technology}
  \city{Hong Kong}
  \country{China}
}

\author{Shuai Ma}
\email{shuai.ma@connect.ust.hk}
\orcid{0000-0002-7658-292X}
\affiliation{%
  \institution{The Hong Kong University of Science and Technology}
  \city{Hong Kong}
  \country{China}
}

\author{Jiehui Luo}
\email{jluobj@connect.ust.hk}
\orcid{0000-0002-1394-4348}
\affiliation{%
  \institution{The Hong Kong University of Science and Technology}
  \city{Hong Kong}
  \country{China}
}

\author{Xiaojuan Ma}
\email{mxj@cse.ust.hk}
\orcid{0000-0002-9847-7784}
\affiliation{%
  \institution{The Hong Kong University of Science and Technology}
  \city{Hong Kong}
  \country{China}
}

\renewcommand{\shortauthors}{Zheng, et al.}

\begin{abstract}
Existing research on human-AI collaborative decision-making focuses mainly on the interaction between AI and individual decision-makers. There is a limited understanding of how AI may perform in group decision-making. This paper presents a wizard-of-oz study in which two participants and an AI form a committee to rank three English essays. One novelty of our study is that we adopt a speculative design by endowing AI equal power to humans in group decision-making. We enable the AI to discuss and vote equally with other human members. We find that although the voice of AI is considered valuable, AI still plays a secondary role in the group because it cannot fully follow the dynamics of the discussion and make progressive contributions. Moreover, the divergent opinions of our participants regarding an ``equal AI'' shed light on the possible future of human-AI relations.
\end{abstract}




\maketitle

\section{Introduction}
Modern Artificial Intelligence (AI) has profoundly changed human decision-making practices. 
We have observed cases where AI replaces humans to make decisions.
From low-stakes decisions such as content moderation~\cite{kim2007youtube} to more critical decisions such as recruitment~\cite{van2019marketing}, AI skyrockets the decision-making efficiency but inevitably causes concerns regarding fairness, accountability, and transparency~\cite{alkhatib2019street, alkhatib2021live, mujtaba2019ethical, shneiderman2022human}.
Consequently, people have argued for the need to have \textit{human-AI collaborative decision-making}.
One typical example is the AI-assisted decision-making system (ADS), in which the predictions from AI serve as suggestions and the final decisions are made by humans~\cite{levy2021assessing, buccinca2021trust, bansal2021does}.

Despite the growing body of research on human-AI collaborative decision-making, far fewer works investigate the collaboration between AI and a group of humans~\cite{lai2021towards}.
\textit{Group decision-making} has a long history in human society. 
It has been praised for its ability to bring together different perspectives on decision-making and to obtain better decisions than individual decision-makers~\cite{saaty2013group, hirokawa1996communication, ivancevich1990organizational}.
\revision{
With the fast development of data infrastructures and the increasingly advanced machine learning (ML) models,
researchers have discussed the potential of AI to improve the group decision efficiency~\cite{trunk2020current}, reduce human bias~\cite{petrin2019corporate}, and offer novel while valuable perspectives to the group~\cite{kleinberg2018human, lai2020chicago}.
Some business organizations have attempted to add AI to their internal meetings for decision-making~\cite{petrin2019corporate, erouglu2022impact}.
For example, Tieto, a Finnish IT company, has announced to ``appoint'' AI to their leadership team~\cite{Tieto}.
Therefore, there is a growing tendency to involve AI in group decision-making~\cite{petrin2019corporate, trunk2020current}. 
We need more knowledge of the human-AI interaction in group scenarios to design AI effectively and responsibly.
}

There are multiple positions that AI may be able to take on in group decision-making.
Some previous studies have shown the benefits of having intelligent agents serve as group coordinators who facilitate group communication but do not directly provide support concerning the actual decisions~\cite{shamekhi2018face, Kim2021Moderator}.
Other works suggest that AI can behave as a decision consultant or assistant, which offers suggestions to human group members only for reference~\cite{yang2019unremarkable}. 
In high-stakes scenarios such as clinical diagnosis, this turns out to be the first choice for security and accountability reasons~\cite{yang2019unremarkable}.
Nevertheless, in not life-or-death decision scenarios, the subordinate position of AI in the group may undermine its potential to contribute to the final decision~\cite{dietvorst2015algorithm}.
While existing research has investigated the equal human-AI relationship in human-robot interaction~\cite{Karthik2021Allocation} and cooperative games~\cite{wehbe2017left}, it is unknown in the context of group decision-making how people will collaborate with and perceive an AI group member that shares the same decision-making power with them.

In this paper, by adopting the frictional research through design (RtD) approach~\cite{pierce2021tension}, we would like to probe a future of \textbf{endowing equal power to AI in human-AI group decision-makings}.
Our goal is not to create an optimal solution for AI service in the future but to discover \textit{frictions} by prototyping and user experimentation to deepen our understanding of the human-AI relationship.
We seek to answer the following research questions (RQs):
\begin{enumerate}
    \item[\textbf{RQ1:}] How do people work with the AI member in group decision-making?
    \item[\textbf{RQ2:}] How do people perceive the participation and contribution of an AI member that is designed to have equal power with them?
\end{enumerate}

We chose \textit{student essays ranking} as the task scenario, since there is an increasing trend of assessing students' work with AI grading systems~\cite{hsu2021attitudes}.
It is also a practical task that involves group decisions in scenarios such as student essay competitions.
In our experimental setting, two teachers and an AI form a committee to score three essays and decide the ranking for awards.
\revision{
The group decision-making process is structured based on the nominal group technique (NGT), which we introduce in detail in Sec.~\ref{sec:GDMProcess}.
Generally, NGT encourages the equal contribution of all group members and is suitable for small groups~\cite{Hagerup1993}.
}
To simulate the AI group member, we adopted a constrained wizard-of-oz approach~\cite{riek2012wizard} and designed the AI in an iterative and human-centered process.
In particular, the predictions and corresponding explanations of the AI member are from a neural-network-based ML model, and its interaction with other group members is controlled by a human wizard.
The AI member participates in the decision-making process on an equal footing with human members from two aspects: 1) the AI member shares ideas and discusses the proposed ranking with other human members by asking as well as answering questions equally, and 2) the AI member can vote for the final group decision just like human members.

To answer the research questions, we conducted an exploratory study with 20 English teachers (ten experimental groups) from middle schools and universities in China.
Our results reveal how human teachers seek opinions from the AI group member and how AI's statement stirred the conversation in the group.
Our participants believed AI could improve the objectivity and fairness of the final group decision (\textit{competent}). Nevertheless, they also suggested that AI cannot really contribute to the discussion due to reasons such as the inability to incorporate human opinions and follow the progress of the discussion (\textit{rigid}).
Moreover, our participants have different opinions on how AI should participate in group decision-making with humans.
In summary, our contributions of this paper are three-fold:
\begin{itemize}
    \item We develop a speculative design of AI that can participate in group decision-making equally with humans via a constrained wizard-of-oz protocol;
    \item We conduct an exploratory study and derive qualitative understandings of how the AI group member collaborates with a group of human experts to make decisions;
    \item We reflect on our design of the AI group member and provide insights on the future directions of studying AI-in-the-group. 
\end{itemize}

\section{Related Work}
\revision{
Our work investigates how humans perceive and interact with an equal AI member in group decision-making.
We relate our work to the research on human-AI collaborative decision-making, the different roles of AI in human-AI teams, and strategies to support group decision-making.
}

\subsection{Human-AI Collaborative Decision-Making}
\label{related-1}
A fundamental motivation to have human-AI collaborative decision-making is AI's and humans' imperfections. Human experts are more trusted by stakeholders~\cite{mou2017media}, but they may need to spend considerable time and effort making decisions.
Sometimes their decisions might also be irrational for reasons like cognitive bias~\cite{samuelson1988status}. On the other hand, although AI can achieve high performance on a wide variety of tasks these days, people still have concerns about: 1) \revision{potential errors from AI that might cause serious results~\cite{alkhatib2019street, Robinette2016, khairat2018reasons}}; 2) fairness and transparency issues~\cite{cheng2022child, kawakami2022improving}; and 3) unawareness of the social and business context~\cite{ehsan2021expanding, zheng2022telling, wang2021brilliant}.

\revision{
One common form of human-AI collaboration on decision-making is that AI works as an assistant to offer suggestions while humans make the final calls~\cite{lai2021towards, bansal2021does, buccinca2021trust}.
Previous works reveal that the suggestions from AI, when highly accurate, can significantly improve humans' performance over a number of tasks~\cite{lai2020chicago, lundberg2018explainable, buccinca2020proxy, shi2022medchemlens}.
Moreover, human monitoring helps reduce unfairness caused by AI decisions~\cite{cheng2022child}.
Nevertheless, new issues, such as humans' inappropriate trust in AI~\cite{zhang2020effect,lai2019human} and the lack of complementary performance~\cite{bansal2021does}, have been found by researchers as obstacles to successful human-AI decision-making.
Previous research suggests that only focusing on improving model performance does not always lead to human-AI team success~\cite{bansal2021most}; instead, human factors, such as trust, engagement and domain expertise, matters~\cite{bansal2021does, wang2020factors, lai2020chicago, nourani2020role, ma2022modeling}.
HCI researchers have proposed various designs of AI systems to better collaborate with humans.
}
For example, Cai et al.~\cite{cai2019human} investigate how experts can refine AI's suggestions and design a series of refinement tools.
Levy et al.~\cite{levy2021assessing} found that, for clinical text annotation, having experts collaborate with AI in two-step can raise experts' self-agency.

We join previous research in exploring how to facilitate better collaboration between AI and experts (i.e., English teachers) on decision-making tasks (i.e., student essay ranking), but focus on group decision-making.
Early research by Foster and Coovert~\cite{10.1145/633292.633378} found that recommendations from a simulated agent, whether true or not, can significantly influence the decision made by a group of students.
More recently, targeting a critical decision-making task, Yang et al.~\cite{yang2019unremarkable} propose an ``unremarkable'' design that positions the model predictions in the corners of slides used in meetings of experts.
They reported that the ``unremarkable'' role of AI could improve decision-making and was well-accepted by clinicians in critical scenarios.
\revision{
Previous research mainly investigates decision supports from ``assisted AI''~\cite{petrin2019corporate}, whereas we explore another design of AI that treats AI as an equal group member to humans.
We hope such a design could inform the benefits and barriers of AI having greater power in decision-making groups.
}

\revision{
\subsection{Different Roles of AI in Human-AI Teams}
\label{related3}
In the past decade, some organizations have tried to allow AI to act in the role of equal group member to humans in the domain of corporate governance.
In 2014, Deep Knowledge, a venture capital firm in Hong Kong, announced to have AI join its board to make investment decisions~\cite{VITAL}. However, later reports revealed that, as Hong Kong does not allow ``non-human entities'' on boards, the AI was only an ``observer'' to avoid incorrect investments~\cite{petrin2019corporate, erouglu2022impact}.
Later in 2016, an IT company named Tieto reported its deployment of AI to ``participate in team meetings'' and ``vote on the business direction''~\cite{Tieto, erouglu2022impact}.
While such business news may be more for marketing purposes, relevant discussion occurs regarding the possible roles of AI in corporations.
Petrin~\cite{petrin2019corporate} postulates three progressive roles of AI on corporate boards: ``assisted AI'', ``advisory AI'' and ``autonomous AI''.
While assisted AI only offers suggestions, advisory AI shares the decision rights with humans, and autonomous AI fully takes the decision rights.
Petrin anticipates the rights of AI would gradually increase and benefit the decision-making to be more objective and more aligned with corporate missions~\cite{petrin2019corporate}.
Concerning the possible future of AI corporate governance, there are also some suspicions.
Eroğlu and Kaya~\cite{erouglu2022impact} suggest the importance of the diversity of human decision-makers on corporate boards considering the potential bias of AI predictions.
Hilb~\cite{hilb2020toward} concerns that AI sharing decision rights challenges the current regulation on accountability and liability~\cite{hilb2020toward}.
Moreover, some researchers explicitly stress that strategic organizational decisions must be made by humans~\cite{trunk2020current}.

Despite the controversy in corporate governance, AI has the potential to be a member of a decision-making team. 
There have been some empirical studies about equal partnerships between humans and AI, but mainly in game-like simulated environments~\cite{schelble2022let, wehbe2017left}. 
When it comes to decision-making tasks in real-world scenarios, 
most research works position AI as an \textit{assistant} that provides suggestions to decision-makers (as discussed in Sec.~\ref{related-1}). 
A few other studies investigate how agents, as a \textit{facilitator}, may aid communication between human members in a decision team.
For example, Kim et al.~\cite{kim2020bot} designed a chatbot to facilitate online group discussion to encourage participation and organize opinions.
Shamekhi et al.~\cite{shamekhi2018face} conducted a wizard-of-oz study in which an embodied agent served as a facilitator in synchronized group decision-making scenarios.
Nevertheless, few HCI studies investigated the effects of having an AI group member with equal decision rights to humans.

To fill this gap, our study tests the idea of having an equal AI member in a structured group decision-making process (i.e., NGT).
Note that the goal of this paper is not to seek equal rights for AI, which would be arguably problematic ethically and legally~\cite{trunk2020current}.
Instead, we treat our design as frictional~\cite{pierce2021tension} and hope that such an empirical study can enrich the discussion of AI in playing certain roles in human-AI teams.
}

\subsection{Support Group Decision Making}

When the decisions to be made are complex and uncertain, people tend to need group (instead of individual) decision-making, even though it takes longer, as the pooled knowledge of the experts in the group makes the final decision better~\cite{ivancevich1990organizational}.
However, different from individual decision-making, group decisions are often influenced by some social behavioral factors such as conformity, namely people tend to change their decisions to follow the majority~\cite{bond2005group}; and social loafing, namely individuals work less hard when they are in groups instead of working alone~\cite{liden2004social}.

Some group techniques are often applied to stimulate better decisions.
For example, the Delphi technique fosters group decisions by collecting anonymous feedback from experts~\cite{linstone1975delphi}; NGT (used in our study) defines a four-stage decision-making procedure to engage every group member~\cite{harvey2012nominal}.
On the other hand, people develop group decision support systems (GDSS) to leverage computers to facilitate the decision-making process.
As early as 1987, DeSanctics and Gallupe~\cite{desanctis1987foundation} envisioned three types of support from GDSS: support generating ideas, support choosing from alternatives, and support communication in the group.
In later ages, many creative GDSS designs have been proposed.
\revision{
For example, Hong et al.~\cite{hong2018collaborative} present a visualization design of group awareness, which improves the group efficiency in reaching a consensus.
Khadpe et al.~\cite{khadpe2021empathosphere} propose a design that can engage group members in others' perspective, which facilitate effective communication within the group.
}

\revision{
While the aforementioned works are focused on providing tools for human group members, we develop an AI group member to participate in the group decision-making and investigate how human members perceive and interact with it.
}
To this end, we design the AI group member with the ability to share ideas, discuss with other members, and vote.
We set up the scenario as synchronous, videoconferencing-based group decision-making, in which group members communicate via voice.
\revision{
As current speech technology cannot perfectly follow real-time human discussions~\cite{li2022recent}, we adopt a wizard-of-oz design, similar to previous research studying human-AI collaboration~\cite{hohenstein2022vero, schelble2022let}, but constrained by the real ability of a deep learning model.
}

\section{Designing AESER: AI Member in Group Decision Making}
To probe the future of having an AI member in group decisions, we applied NGT~\cite{Hagerup1993} to design a human-AI collaborative group decision-making process under the task of ranking three student essays.
To this end, we designed and developed an intelligent agent named \textbf{AESER} (Automated Essay ScorER) that can score individual student essays and work with human teachers following NGT to determine the final ranking as a team. 

To develop AESER, we first designed a neural-network-based ML model, which determines the scores AESER would assign to essays and the explanations AESER would provide.
Second, to enable AESER to interact with human group members, we developed a constrained Wizard-of-Oz protocol~\cite{riek2012wizard, shamekhi2018face}.
The wizard can only control AESER \textit{asking} or \textit{answering} questions based on a limited number of pre-defined rules.
By constraining the available actions of the wizard, we make the behavior of AESER close to what today's AI technology can really support and make our study more realistic.

In the following sections, we first introduce the task and the group decision-making process.
Then, we detail the design of AESER, including the iterative design process, its backend ML model, and the wizard-of-oz protocol.

\subsection{The Task: Student Essay Ranking}
We choose the task of ranking three student essays as a representative case for group decision-making.
Our task choice is motivated by several considerations.
First, the essay ranking task involves activities such as comparing candidates and making sense of the pros and cons of each candidate, which is similar to tasks used in previous HCI studies on group decision-making~\cite{farnham2000structured, shamekhi2018face, khadpe2021empathosphere}.
Second, such a task is closely related to real-world scenarios, for example, student essay competitions.
Moreover, one critical subtask under essay ranking is essay scoring, which has been fully studied in ML community~\cite{taghipour2016neural,kumar-etal-2022-many}.
Similar ML-based scoring tools have been deployed in the real world~\cite{hsu2021attitudes}.
Thus, it is more conceivable and realistic to make AI a member of the decision group for such a task.

We use the Automated Student Assessment Prize (ASAP) dataset~\footnote{\url{https://www.kaggle.com/competitions/asap-aes/data}} for our task.
The ASAP dataset comprises eight essay sets corresponding to different essay prompts.
The first six essay sets in ASAP are labeled only for the overall essay scores, whereas essay sets 7 and 8 have additional trait score labels (e.g., language style score).
We believe if AI could provide trait scores, more group discussions could be sparked.
We thus choose \textit{essay set 7} which have four traits -- \textit{Content}, \textit{Organization}, \textit{Style} and \textit{Convention} -- as our task dataset, of which essays are written by US high school students from grade 7.
We did not choose essay set 8 with six traits because it might cause too much workload for participants in our experiment.
The prompt of \textit{essay set 7} asks students to write a story about patience.

\subsection{Group Decision-Making Procedure}
\label{sec:GDMProcess}
Although our goal is to make AI an equal group member to humans, it is still challenging to build an intelligent agent that can join \textit{free-form group discussions}.
Considering the technical difficulty, we leverage NGT -- a widely adopted technique in areas such as health, education, and social service -- to make the group decision-making process structured and controllable~\cite{ivancevich1990organizational}.
NGT requires group members to first generate their own ideas silently and then share and discuss their ideas with other members~\cite{harvey2012nominal, potter2004nominal}.
The final decision is often based on a voting round at the end of the discussion.
NGT is particularly suitable for small group processes and encourages even contribution~\cite{harvey2012nominal, potter2004nominal}.
To be more specific, the group decision process in our experiment contains the following four steps (illustrated in Fig.~\ref{fig:procedure}):

\begin{enumerate}
    \item[Step1] \textbf{Silent essay review.} Every group member has 15 minutes to read through the given essays independently. They need to review the essays, provide trait scores, and propose an essay ranking with clear arguments.
    \item[Step2] \textbf{Essays ranking sharing.} In random order, each group member will share their essay rankings and arguments with other members within three minutes. 
    \item[Step3] \textbf{Group discussion.} This step is for group members to clarify their rankings through back-and-forth question-answering. All group members together have 15 minutes to ask questions to each other regarding any confusion about or differences between their proposed scores.
    \item[Step4] \textbf{Final voting.} Group members complete a questionnaire to suggest their final rankings separately. The final group decision depends on the majority opinion. For example, for first place in the ranking, if two group member vote for essay A and one group member votes for B, essay A would be placed first~\footnote{There could be cases where there is no majority vote. For example, the three group members vote for different essays for the same ranking position. Although we did not encounter such situations in our studies, in original design, we would ask the group to go back to \textit{Step 3} for further discussion and then vote again. If the second round of voting still cannot reach a consensus, the group decision-making process would end with no result.}. 
\end{enumerate}

\subsection{The Iterative Design Process}
We ran two rounds of pilot studies to improve the design of AESER iteratively.
For each round of studies, we recruited two groups of college students (two human members per team, a total of four females and four males) to collaborate with AESER to go through the group decision-making process.
Participants in each study were invited to a 15-minute follow-up interview about their experience interacting with AESER.
Two authors analyzed all transcripts of the interviews using inductive thematic analysis~\cite{guest2011applied}.
Conflicts were resolved through constant communication.
Key design choices resulting from the pilot studies are as follows:


\begin{itemize}
	\item \textbf{AESER needs a set of pre-compiled answers regarding questions on its global behavior}. Initially, AESER only answer questions regarding its predictions. However, participants also wanted to know some global information such as AESER's capability and the model behind it.
	In response, we prepare a set of answers for AESER in its final implementation based on the question bank of user-centered explainable AI (XAI)~\cite{liao2020questioning}.


    \item \textbf{AESER will share its idea in an uncertain tone when the backend ML model makes predictions with low confidence}. The participants in our pilot studies found that when AESER explains its decisions, ``\textit{it is very aggressive, directly providing a conclusion}'', which sounds like a ``\textit{reference answer}''. They consider such behavior makes AESER less like an equal group member. Participants suggested AESER should express its uncertainty whenever it is unsure.

    \item \textbf{The questions that AESER asks are generated automatically, and the wizard only controls when AESER asks them}. In our pilot studies, the wizard needs to manually come up with the questions for AESER to pose, which turns out to be a heavy workload and may make AESER's behavior less consistent. In our final implementation, the questions AESER raises are all automatically computed based on the difference between the scores it gives and the scores from the other members (see Sec.~\ref{sec:wizard}).
\end{itemize}


\begin{figure*}
    \centering
    \includegraphics[width=\linewidth]{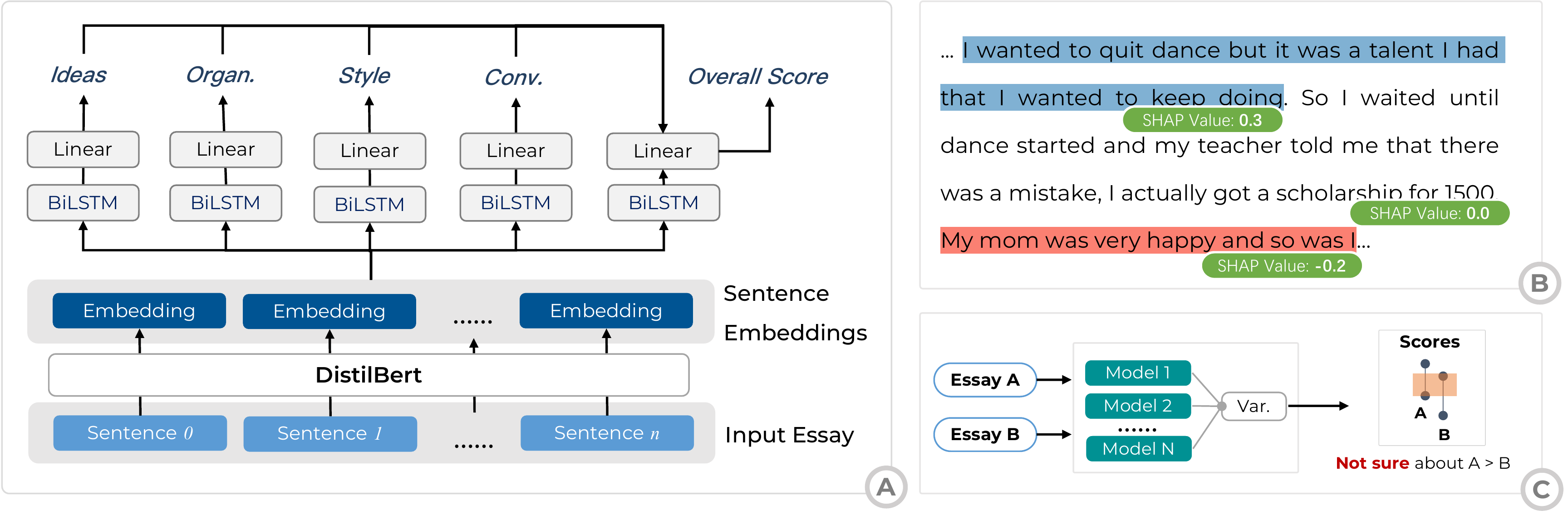}
    \caption{(A) Architecture of the ML model, which computes five scores (four trait scores and one overall score) from an input student essay; (B) AESER explains its predictions at sentence-level; (C) AESER expresses uncertainty when the confidence interval of the scores of two essays overlap.}
    \Description{
    (A) is an ML model architecture illustration. The input of the model is student essays. The input essay is represented as a sequence of sentences, which are sent to a DistilBert module to output a sequence of sentence embeddings. The embeddings are then sent to five scoring modules. Four are for trait scores, and one is for the overall score. The score modules are composed of a BiLSTM layer and a Linear layer. Especially the Linear layer of the overall score module also takes the four trait scores as input. (B) contains a paragraph of text in which two sentences are highlighted and tagged with their SHAP values (0.3 and -0.2, respectively). in (C), two essays are fed into multiple models, and the variance of the models' given scores of each essay is computed. If the confidence intervals of the scores of essays, which are based on the computed variance, overlap, AESER is not sure about the comparision of the two essays.
    }
    \label{fig:ml_model}
\end{figure*}

\subsection{Machine Learning Model}
\label{sec:model}

Our ML model was trained for a subtask of \textit{student essay ranking} -- automatic essay grading (AEG), which has been well studied in ML communities~\cite{kumar-etal-2022-many, taghipour2016neural}.
AESER simply determines the ranking of several candidate essays by sorting them based on the scores produced by the model. 

\subsubsection{Model Architecture and its Performance}
\label{sec:model-architecture}

We designed and trained an ML model with similar architecture to the AEG model in Kumar et al.~\cite{kumar-etal-2022-many}.
\revision{
As shown in Figure~\ref{fig:ml_model}A, by adopting a multi-task learning approach, the model can compute both \textit{the overall score} and four \textit{trait scores} for each essay.

Specifically, each input essay is converted into an ordered sequence of sentences to feed into the model.
We use a pre-trained DistilBert layer~\cite{sanh2019distilbert} to transform the original sentences into embeddings.
Next, similar to Kumar et al.~\cite{kumar-etal-2022-many}, for each type of score, the sentence embeddings go through a BiLSTM layer~\cite{graves2013speech} to aggregate the information of all sentences and form an essay representation.
The essay representation is sent to linear layers to output the trait scores.
For the overall score, the essay representation is concatenated with the trait scores and then sent to a linear layer for score computation.
The model is optimized based on the mean squared error loss and trained end-to-end.

We applied a 60\%, 20\%, 20\% split of training, validation, and test sets for the data in \textit{essay set 7} and evaluated the model on the test set using quadratic weighted Kappa (QWK), similar to previous research~\cite{taghipour2016neural, kumar-etal-2022-many}.
QWK measures the agreement between the models' output scores and the human-assigned scores provided in the dataset. }
The results show that the model achieves 0.815 QWK score, slightly better than Kumar et al.'s model~\cite{kumar-etal-2022-many} and achieves state-of-the-art performance~\cite{taghipour2016neural}.
The model also performs well in predicting trait scores, with QWK scores of 0.780, 0.713, 0.766, 0.748 for trait \textit{Content}, \textit{Organization}, \textit{Style} and \textit{Conventions}, respectively.



\subsubsection{Explaining the ML Model}
\label{xai-design}
Explanation is essential for the AI group member to join group discussions.
We conducted a survey study with English teachers asking how they usually explain their scoring to other teachers.
We presented them with a range of explanation methods derived from existing XAI literature as options~\cite{liao2021human, Guidotti2018XAIsurvey}.
Informants were invited to select the one(s) they found helpful.
\revision{
We received 42 responses from informants (31 self-identified as females and 11 as males) with essay-scoring experience and at least one year of English teaching experience.
Their mean age is 27.0 years (SD = 6.3 years).
}

\begin{table*}[htbp]
	\centering
	\begin{tabular}{lcc}\toprule
		Survey options & Related XAI method & Voting rate \\\midrule
		Keywords & Word-level feature-based methods & 44.2\% \\
		Key sentence & Sentence-level feature-based methods & 69.2\% \\
		Sample essays with similar scores & Example-based explanations & 36.5\% \\ 
		Modification suggestions & Counterfactual inspection & 25.0\% \\ \bottomrule
	\end{tabular}
	\caption{The feedback from English teachers on the preference of explaining their student essay scorings}
	\label{tab:survey}
\end{table*}

As the survey results summarized in Table \ref{tab:survey} suggested, explaining the essay scores by showing the corresponding ``key sentences'' received the most votes from our respondents.
We thus decided to use the sentence-level feature-based XAI method. 
In particular, we adopted Shapley additive explanations (SHAP)~\cite{NIPS2017_7062}.
With an ML predictor, SHAP computes a value (called SHAP value) for each sentence of any input essay (Fig.~\ref{fig:ml_model}B).
The SHAP value, which can be positive or negative, represents the contribution of the sentence to the prediction. 
Using SHAP, we can identify the key sentences that play a part in the ML model's prediction of the overall essay score and the four trait scores, respectively. 
\revision{
AESER's answers to other group members' questions can be composed using these sentences.
For example, if a human member asks, ``\textit{why do you give a higher score to essay A on its organization than essay B}'', AESER would use the sentences in essay A whose SHAP values are the most positive in predicting the ``organization'' score to explain the high score of essay A; sentences in essay B whose SHAP values are the most negative would also be presented for justifying the weakness of essay B.
We provide more details on how AESER answers questions in Sec.\ref{sec:answerquestions}.
}

\subsubsection{Evaluating the Uncertainty of Model Predictions}

We applied \textit{deep ensemble}, one of the most effective uncertainty estimation methods for neural networks~\cite{ovadia2019can, lakshminarayanan2017simple}.
Specifically, we trained five models with the same architecture introduced in Sec.~\ref{sec:model-architecture} but with different parameter initialization.
For each input essay, a confidence interval is derived by computing the standard variance of the model's predictions.
When AESER compares two essays on any score, if the confidence intervals of these two essays overlap, AESER would indicate that it is unsure in comparing these two essays on that particular type of score (Fig.~\ref{fig:ml_model}C).


\subsection{The Wizard-of-Oz Protocol}
\label{sec:wizard}
\begin{figure*}
    \centering
    \includegraphics[width=0.95\linewidth]{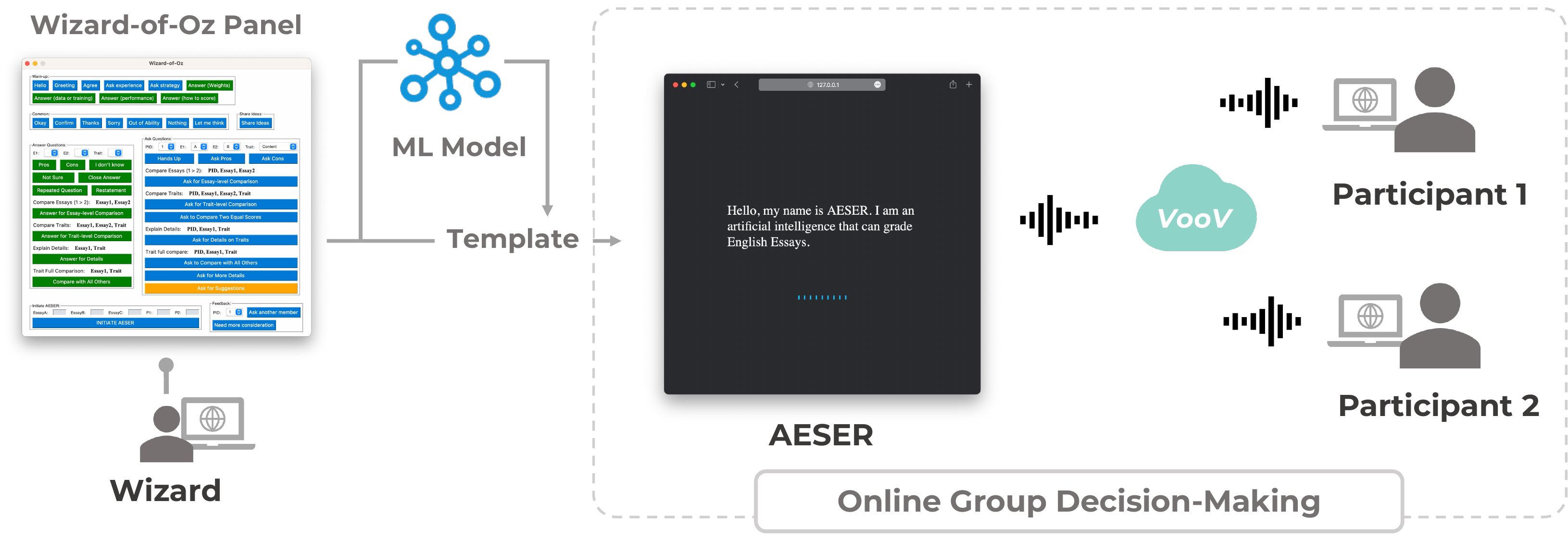}
    \caption{Illustration of the Wizard-of-Oz study. Two participants and AESER make decisions together by attending an online meeting. A human wizard controls AESER through a control panel. All wizard's instructions related to AESER's given scores will turn into a speech by 1) weaving a text by extracting necessary information from the ML model and 2) using an online text-to-speech service. Other instructions would revoke a template response. }
    \Description{A wizard is controlling a wizard-of-oz panel, which control the ML model and the template selection of AESER's responses. AESER is collaborating with two participants in the online group decision-making over the VooV platform.}
    \label{fig:wizard}
\end{figure*}

We simulated how AESER, if fully automated, would interact with humans in a realistic setting by developing a constrained wizard-of-oz protocol~\cite{riek2012wizard}.
Similar to prior works~\cite{shamekhi2018face, robe2022designing, kuttal2021trade}, conversation-related functions, such as speech recognition, natural language understanding, and response generation, are enabled by a human wizard by following well-calibrated protocol (e.g., Tab.~\ref{tab:answers}).
The wizard instructs AESER to produce an utterance either from a pre-defined template text (e.g., ``\textit{I don't know}'') or by assembling the model's information into a template with open slots (Fig.~\ref{fig:wizard}).
We use an online Text-to-Speech service~\footnote{\url{https://cloud.tencent.com/product/tts}} to generate the voice of AESER from the output text scripts.
All text and speech are in Chinese.
We translated them into English in this paper.

In each group decision-making session, the researcher who played the role of the wizard shared his laptop's screen and audio in the online meeting.
All group members can hear the speech from AESER and see the corresponding text transcription from the shared screen (Fig.~\ref{fig:wizard}~right).
To them, the wizard is merely an observer of the group process.

In the following subsections, we introduce several essential features of AESER enabled by the wizard-of-oz protocol.
These features, together with \textit{voting} in step 1 and step 4 of NGT, allow AESER to act like a group member to participate equally with human teachers in group decision-making.

\subsubsection{Share the Ranking}
\begin{figure*}
    \centering
    \includegraphics[width=0.95\linewidth]{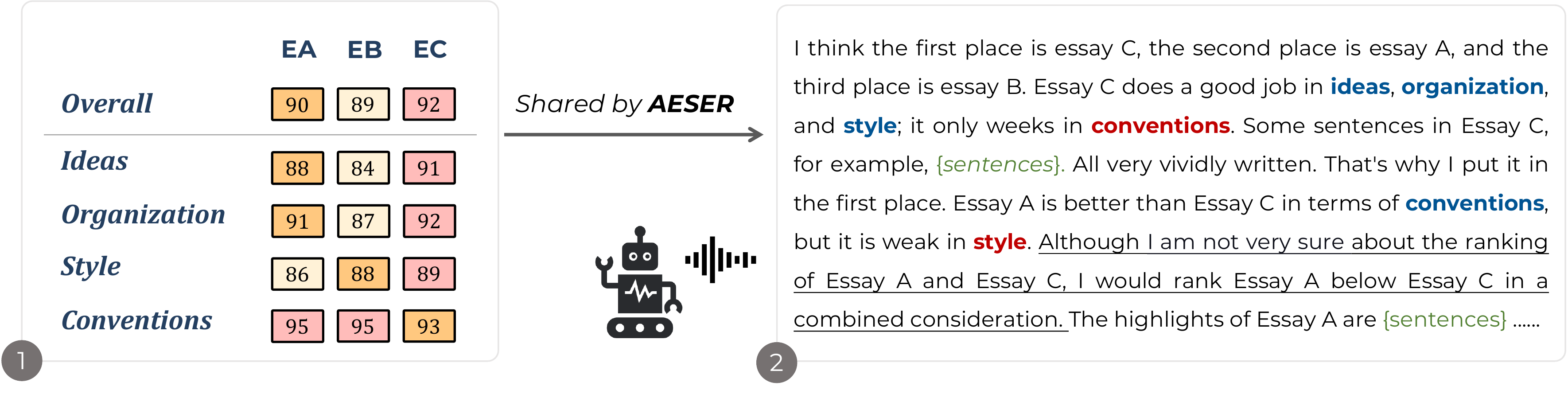}
    \caption{Demonstrate how AESER shares its ranking: (1) The prediction of three essays' overall scores and trait scores. The shadow encodes the ranking level. The resulting ranking is $EC > EA > EB$; (2) The corresponding text (partial) shared by AESER in step 2 of NGT. $\{sentences\}$ refers to explanations generated by SHAP for the overall score of the specific essay.}
    \Description{(1) is a scoring matrix. The overall, ideas, organization, style and converntions scores of the EA, EB and EC are (90, 88, 91, 86, 95), (89, 84, 87, 88, 95), (92, 91, 92, 89, 92), respectively. The scoring matrix is shared by AESER as ``I think the first place is essay C, the second place is essay A, and the third place is essay B. Essay C does a good job in ideas, organization, and style; it only weeks in conventions. Some sentences in Essay C, for example, {sentences}. All very vividly written. That's why I put it in the first place. Essay A is better than Essay C in terms of conventions, but it is weak in style. Although I am not very sure about the ranking of Essay A and Essay C, I would rank Essay A below Essay C in a combined consideration. The highlights of Essay A are {sentences} ...''.}
    \label{fig:weave_idea}
\end{figure*}

In \textit{Essay ranking sharing} (Step 2 of NGT), the wizard instructs AESER to share its ranking of the three essays.
Fig.~\ref{fig:weave_idea} presents an example.
AESER first states its proposed ranking, which is based on the \textit{the overall scores}.
Then, AESER presents its arguments for the ranking of the three essays one by one.
It highlights the strengths (\textbf{blue text} in Fig.~\ref{fig:weave_idea}) and weaknesses (\textbf{red text}), if any, of each essay compared to the other candidates.
For individual essays, AESER quotes the sentences with either the highest positive SHAP value or the lowest negative SHAP value (if exists) based on the model's prediction of overall scores as evidence for its argument.
If AESER is not sure about the relative positions of two essays (i.e., the confidence intervals of the predictions overlap), it would explicitly indicate the uncertainty (underlined text in Fig.~\ref{fig:weave_idea}).

\subsubsection{Answer Questions about Given Scores}
\label{sec:answerquestions}
\begin{table*}[htbp]
 \centering
 \begin{tabular}{llc}\toprule
    Question Type & Question & Template of Answering \\\midrule
    \parbox{2.75cm}{Overall Comparison} & \parbox{3.5cm}{Why do you think \textit{\{EX\}} is better than \textit{\{EY\}}?} & \parbox{8cm}{I think \textit{<EX>} is better than \textit{<EY>} mainly in \textit{<Traits>}. For example, on \textit{<trait>}, some sentences of \textit{<EX>} is well written: \textit{<sentences>}; on the other hand, some sentences of \textit{<EY>} appear to be poorly done in terms of \textit{<trait>}, such as: \textit{<sentences>}} \\ \cr
    \parbox{2.75cm}{Trait-level Comparison} & \parbox{3.5cm}{Why do you think \textit{\{EX\}} is better than \textit{\{EY\}} in terms of \textit{\{trait\}}?} & \parbox{8cm}{The main requirement of \textit{<trait>} is \textit{<trait descriptions>}.  I think a few sentences of \textit{<EX>} show that it's doing a good job of \textit{<trait>}. For example, \textit{<sentences>}. \textit{<EY>} has some nice sentences, for example, \textit{<sentences>}. However, there are also some unreasonable sentences in \textit{<EY>}: \textit{<sentences>}} \\ \cr
    \parbox{2.75cm}{Details on Trait-level Scoring} & \parbox{3.5cm}{How did you grade \textit{\{essay\}}? in terms of \textit{\{trait\}}?} & \parbox{8cm}{The main requirement for \textit{<trait>} is \textit{<trait descriptions>}. I think \textit{<essay>} has a few sentences that exemplify how well it does this. For example \textit{<sentences>}. However, there are also some unreasonable sentences such as \textit{<sentences>}} \\
    \bottomrule
 \end{tabular}
 \caption{A subset of template answers for questions related to AESER's given scores that might be raised by human teachers. \textit{\{Texts\}} in braces are parameters that need to be set by the wizard based on the real questions; \textit{<Texts>} in angle brackets are fields that would be automatically filled based on the question type, provided parameters, and output information from the ML model. }
 \label{tab:answers}
\end{table*}

During group discussions (step 3 in NGT), AESER is expected to participate and answer any questions the human members raise regarding its given scores.
Table~\ref{tab:answers} provides examples explaining how AESER handles questions.
The wizard is responsible for recognizing \textbf{the question type} and entering necessary \textbf{parameters} (e.g., which essay the human teacher mentions) on a wizard-of-oz panel (Fig.~\ref{fig:wizard} left).
Based on the question type, AESER would retrieve the corresponding answer template and fill in the open slots, if applicable, with the parameters entered by the wizard.
Moreover, any uncertainty regarding AESER's predictions would be explicitly delivered in its responses.
For example, if a teacher asks AESER to collate two essays in terms of their styles while AESER's predictions of the style scores overlap, AESER would indicate, ``\textit{I am not very confident about the comparison of the two essays on style}''.
Note that if the wizard did not understand a question clearly, he would command AESER to request the human teacher to restate it, (``\textit{Sorry, I didn't catch your question. Could you please repeat it?}'').

\subsubsection{Ask Questions}

If only human teachers ask AESER questions but not vice versa, we can hardly say that AESER and other teachers participate equally in group decision-making.
Hence, every question we have prepared AESER to \textit{answer} should be a plausible question for AESER to \textit{ask}.
Among these questions, AESER should focus particularly on inspecting and resolving the decision contradictions it has with other human teachers.

\begin{figure*}
    \centering
    \includegraphics[width=\linewidth]{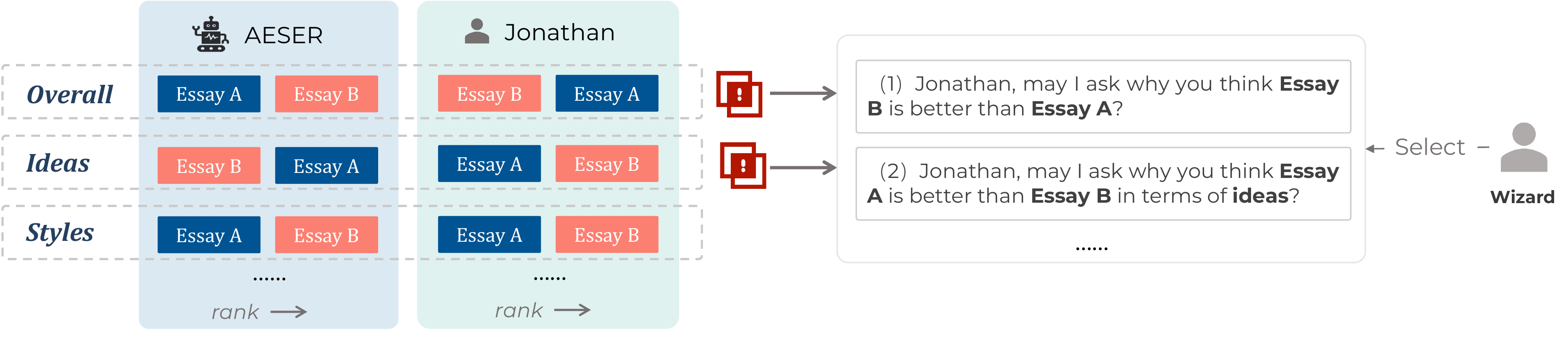}
    \caption{An example of how AESER asks questions. For each pair of essays (e.g., Essay A and B) and each type of score, the contradictions between AESER and another teacher (e.g., Jonathan) are detected. Each contradiction introduces a question for the human teacher.}
    \Description{For the relative rankings of essay A and B in terms of the overall, ideas, and styles scores, AESER gives (A, B), (B, A), and (A, B), respectively. Jonathan gives (B, A), (A, B), and (A, B), respectively. The backend algorithm detects the ranking conflicts and prepares questions such as ``Jonathan, may I ask why you think Essay B is better than Essay A?'' and ``Jonathan, may I ask why you think Essay A is better than Essay B in terms of ideas?'' for the wizard to select from.}
    \label{fig:asking}
\end{figure*}

Inspired by our pilot study results, we implemented an algorithm that can detect the ranking discrepancies that arise from the scores (overall and for individual aspects) given by AESER and those of other human teachers.
Differences found result in a compiled list of questions for AESER to pose.
Figure~\ref{fig:asking} presents an example of AESER's asking process.
With the set of pre-generated, discrepancy-related questions, the wizard's task is to choose which one to ask and when.
Basically, when the whole group is silent for approximately 5 seconds, the wizard will select a question that has not been covered in the previous group discussion for AESER to raise.
\revision{
As AESER's questions involve a switch of topics, we desire AESER's questions to occur when human teachers complete a topic discussion and have no more comments.
Thus, we set 5-second silence, slightly longer than the common silence that occurs in human-human communication~\cite{curhan2022silence}, to be the threshold.
}
In the example of Fig.~\ref{fig:asking}, if the two human teachers only have discussed the \textit{styles} and stop the conversation for a short while, the wizard would instruct AESER to ask a question (2) that concerns a conflicted ranking over \textit{ideas}.
This mechanism is reasonable for implementation and makes our study more realistic.

\subsubsection{Social Interactions}

Inspired by previous research on human-agent interaction~\cite{robe2022designing, kuttal2021trade, shamekhi2018face}, we designed AESER to have active social interactions with other human teachers, including but not limited to greeting, self-introduction (``\textit{I am AESER, an AI that can grade English essays...}''), backchanneling (``\textit{I see}''), and expressing gratitude (``\textit{Thank you}'').

\subsubsection{Refine Predicted Scores}
\label{sec:refineScores}

In theory, if there are conflicts in a group, the optimal solution is performing collaborative problem-solving to merge insights and address all parties' concerns~\cite{ivancevich1990organizational}.
However, few computational methods currently exist, allowing AI to incorporate its thoughts with multiple humans' opinions.
Therefore, we adopt a simple rule-based approach to adjusting AESER's ranking to avoid AESER being too ``stubborn'', especially when its predictions are strongly disagreed by others. 
Specifically, if AESER is unsure about the ranking of two essays and other humans hold different opinions about its proposal, AESER would be instructed to ask something like ``\textit{Since I'm not very sure about the comparison between Essay A and B, do you think it would be better if I rank Essay A before B?}'' at the end of the group discussion.
If all human teachers give an affirmative answer, AESER will change the ranking in its final voting (step 4 in NGT).


\section{Experiment Design}
We conducted an exploratory user study to investigate how English teachers interact with AESER in a group decision-making task and how they perceive its participation.
The study received institutional IRB approval and was conducted online through videoconferencing.

\subsection{Participants}
\begin{table*}[htbp]
 \centering
 \begin{tabular}{llccccc}\toprule
    Group & Participant & Gender & Age & Teaching Experience & Scoring Experience & Discussion Experience \\\midrule
    G1 & P1 & F & [26, 30] & 1 year & More than 10 times & More than 3 times \\
       & P2 & F & [20, 25] & 1 year & Less than 3 times & None \\
    G2 & P3 & M & [26, 30] & 2 years & 3--10 times & 1--3 times \\
       & P4 & F & [20, 25] & 3 years & More than 10 times & More than 3 times \\
    G3 & P5 & F & [20, 25] &1 year & More than 10 times & More than 3 times \\
       & P6 & F & [20, 25] &1 year & More than 10 times & 1--3 times \\
    G4 & P7 & M & [20, 25] & 1 year & 1--3 times & 1--3 times \\
       & P8 & F & [26, 30] & 1 year & 3--10 times & None \\
    G5 & P9 & F & [51, 60] & Over 20 years & More than 10 times & More than 3 times \\
       & P10 & F & [31, 40] & 18 years & More than 10 times & More than 3 times \\
    G6 & P11 & F & [26, 30] & 3 years & More than 10 times & More than 3 times \\
       & P12 & F & [26, 30] & 4 years & 3--10 times & 1--3 times \\
    G7 & P13 & M & [26, 30] & 4 years & More than 10 times & More than 3 times \\
       & P14 & F & [26, 30] & 4 years & More than 10 times & More than 3 times \\
    G8 & P15 & F & [26, 30] & 1 year & 3--10 times & 1--3 times \\
       & P16 & F & [20, 25] & 1 year & 1--3 times & 1--3 times \\
    G9 & P17 & F & [51, 60] & Over 20 years & More than 10 times & More than 3 times \\
       & P18 & M & [41, 50] & Over 20 years & More than 10 times & More than 3 times \\
    G10 & P19 & F & [20, 35] & 1 year & 3--10 times & 1--3 times \\
       & P20 & F & [31, 40] & 5 years & More than 10 times & More than 3 times \\
    \bottomrule
 \end{tabular}
 \caption{\revision{Demographics of participants (Female = F, Male = M). The ages of participants were collected in forms of ranges. The listed experiences refer to their experience of being an English teacher, scoring student English essays, and discussing with other English teachers.}}
 \label{tab:participants}
\end{table*}
\revision{
By contacting informants who filled out our previous survey (Section~\ref{xai-design}), four were available to participate in our user study (P2, P3, P11, and P15 in Tab.~\ref{tab:participants}).
We recruited more participants through our social network and by snowball sampling.
In total, we had 20 participants (16 females and 4 males).
}
All but two participants are English teachers from middle schools in China.
The other two (P17 and P18) are English instructors from a university in China.
The background information of all participants is presented in Tab.~\ref{tab:participants}.

We acknowledge that the student essays used in our study, as well as those fed to AESER for training, were written by students from the United States, while our participants were English teachers from China.
Nevertheless, our participants suggested that the scoring criteria specified in the dataset are very similar to what they used to grade students' essays in practice.
Moreover, our research focuses on how human teachers collaborate with AESER and other human teachers instead of their scoring performance. 
We also recognize the unbalanced gender ratio of our participants, though we had tried to recruit male English teachers through word of mouth. 
While we did not find official statistics regarding the gender distribution of English teachers in China, many reports and online discussions point to the low number of male English teachers in China~\footnote{\url{https://www.zhihu.com/question/52691128}}\footnote{\url{https://new.qq.com/omn/20210309/20210309A04T3S00.html}}.
That is to say, the skewed gender percentage in our study reflects the actual situation to a good extent, but it may limit the generalizability of our findings. 
\revision{
Besides, one may find that the levels of teaching experience were mostly similar for participants within each of our ten experimental groups.
The pairing of participants was only based on the available time slots participants signed up, and we did not deliberately pair them based on other conditions.
We acknowledge that such a pairing result would be a limitation of our study. Future studies should test more diverse grouping compositions.}

\subsection{Experimental Setup}
The three essays used in our study were randomly selected from the testing set of the ASAP dataset and satisfied the following constraints.
First, the three essays are closed in their labeled scores, so it would not be too easy for participants to determine the ranking.
Second, the predicted scores from AESER differ across the three essays, so a ranking can be derived from the predictions.
For each study, the three chosen essays were packed into a PDF file in a random order, together with the detailed scoring criteria.
The PDF file was then distributed to the participants before the start of the main group decision task. 

For all study sessions, one researcher served the wizard's role to ensure the behavior consistency of AESER.
\revision{The wizard turned off his camera and muted himself during the formal experiment, so the participants would not see the actions of the wizard from their points of view.}
Another researcher worked as the coordinator for the group decision-making process.
The coordinator reminded participants of the time limits in each step, guided the group to proceed to the next step once the prior one had finished, and announced the voting results.
He did not involve in the discussion or interfered with participants' decision-making.

\subsection{Procedure}

\begin{figure*}
    \centering
    \includegraphics[width=0.8\linewidth]{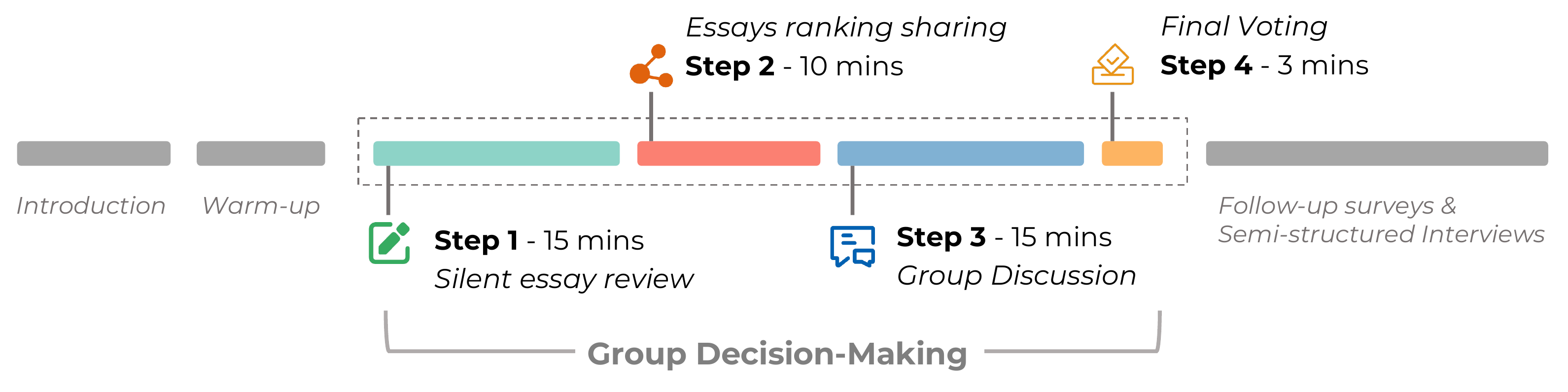}
    \caption{The detailed procedure of our exploratory experiment}
    \Description{There are four stages: introduction, warn-up, group decision-making, and follow-up surveys & semi-structured interviews. For group decision-making, there are four steps: step 1 is 15 mins of silent essay review, step 2 is 10 mins of essay ranking sharing, step 3 is 15 mins of group discussion, step 4 is 3 mins of final voting.}
    \label{fig:procedure}
\end{figure*}

All the studies were conducted over VooV meetings~\footnote{\url{https://voovmeeting.com/}}, an online videoconferencing tool that is similar to Zoom but is more commonly used by our participants (English teachers in China).
We obtained every participant's consent prior to the experiment. Each study session lasted about 90 minutes and included the following four stages (as in Fig.~\ref{fig:procedure}):
1) One researcher introduced the task and the detailed NGT process for group  decision-making.
2) Then, the two participating teachers and AESER had a maximum of 10 minutes to greet and get familiar with each other.
3) Next, the three-member group went through decision-making procedure introduced in Sec.~\ref{sec:GDMProcess}.
\revision{
Tasks that need to be completed, as well as the order of the tasks, are exactly the same for human teachers and AESER at this stage.
}
4) Finally, two teachers filled out a post-study survey. Upon completion, we conducted a 40-minutes semi-structured interview with them.
At the end of each study, we also revealed how AESER work, i.e., the scores and explanations are based on an ML model, but a human wizard triggers the interactions. 
We present each participant a \$30 gift card as a token of appreciation for their time and effort.

\subsection{Measures}
We audio-recorded all sessions with participants' oral consent.
Four types of data were collected from the study:
\begin{itemize}
    \item Participants' scoring of the three essays in step 1 and step 4 of NGT indicated in their questionnaire responses. Participants need to fill in the trait scores of each essay, give their rankings, and provide their confidence in and rationale for their rankings.
    \item Participants' sharing of thoughts (step 2 of NGT) and their discussion with others and with AESER (step 3 of NGT) during the group decision-making process.
    \item Post-study survey results.
    \item Semi-structured interview responses.
\end{itemize}

The post-study survey contains a series of quantitative measures framed as seven-point Likert scale questions, each asking for participants' agreement on one statement.
\revision{
To understand participants' perception of AESER's role in the group, we included statements such as \textit{``AESER works like our assistant''} and \textit{``AESER has an equal position with human members''}.
As a group member, AESER might exert force on other members' decision-making.
Thus, we asked participants to rate how much they perceived AESER to be ``\textit{confident}'' and ``\textit{powerful}''~\cite{shamekhi2018face}.
In addition, we asked about their attitudes on the future adoption~\cite{zheng2022telling, wang2022documentation} of AI in group decision-making.

To understand participants' perception of their social connection with AESER, we collected their ratings on the perceived rapport~\cite{zhao2014towards} (e.g., ``\textit{I get along well with AESER}''), cohesion~\cite{salas2015measuring} (e.g., ``''\textit{The decision-making would be better without AESER}), and trust~\cite{shamekhi2018face} in AESER (e.g., ``\textit{I found AESER trustworthy}'').
}
Besides, we wonder whether our participants have different perceptions of their human peers and AESER. 
Thus, we asked the same set of questions above but substituted ``AESER'' with ``the other human teacher''.


In the semi-structured interviews, we want to know participants' experiences and reflections on the group process.
We asked open-ended questions such as ``\textit{Did AESER's sharing influence your decision?}'', ``\textit{What impact did AESER's questions have on the group discussion?}'' and ``\textit{How did the fact that AESER shared voting rights influence your judgment?}''.
To better put the participants' responses in the context of their previous experiences with group decisions, we asked questions such as ``\textit{What were your past experiences with group decision making?}'', ``\textit{How differently did you behave from your prior participation in all-human group decisions?}'', and ``\textit{Would you like to have AI involved in your group decisions in the future?}''.

\subsection{Data Analysis}
All audio recordings were transcribed into text automatically by the built-in tool of the VooV meeting software.
We manually checked the transcripts to ensure fidelity.
As all the studies were Chinese, we conducted the qualitative coding process and identified themes in Chinese too.
We translate the themes and quotes into English in this paper.
Such a process is similar to previous qualitative studies conducted in Chinese (e.g.,~\cite{wang2021brilliant}).
We applied inductive thematic analysis~\cite{guest2011applied, braun2006using} to analyze the transcripts qualitatively.
Two researchers, who are fluent both in Chinese as well as English, and worked in the user study as the coordinator and the wizard behind AESER, were responsible for coding the data.
They first reviewed all recordings several times to get familiar with and engage in the data.
Then, they independently coded the data of all study sessions in the same order using ATLAS.ti~\footnote{\url{https://atlasti.com/}} and frequently met during the analysis to discuss discrepancies.
After key themes emerged from the data, our research team had several rounds of discussions to refine the results. The entire analysis process lasted several weeks.
Our findings were triangulated with the quantitative results of the post-study surveys.

\section{Results}
\begin{figure*}
    \centering
    \includegraphics[width=0.9\linewidth]{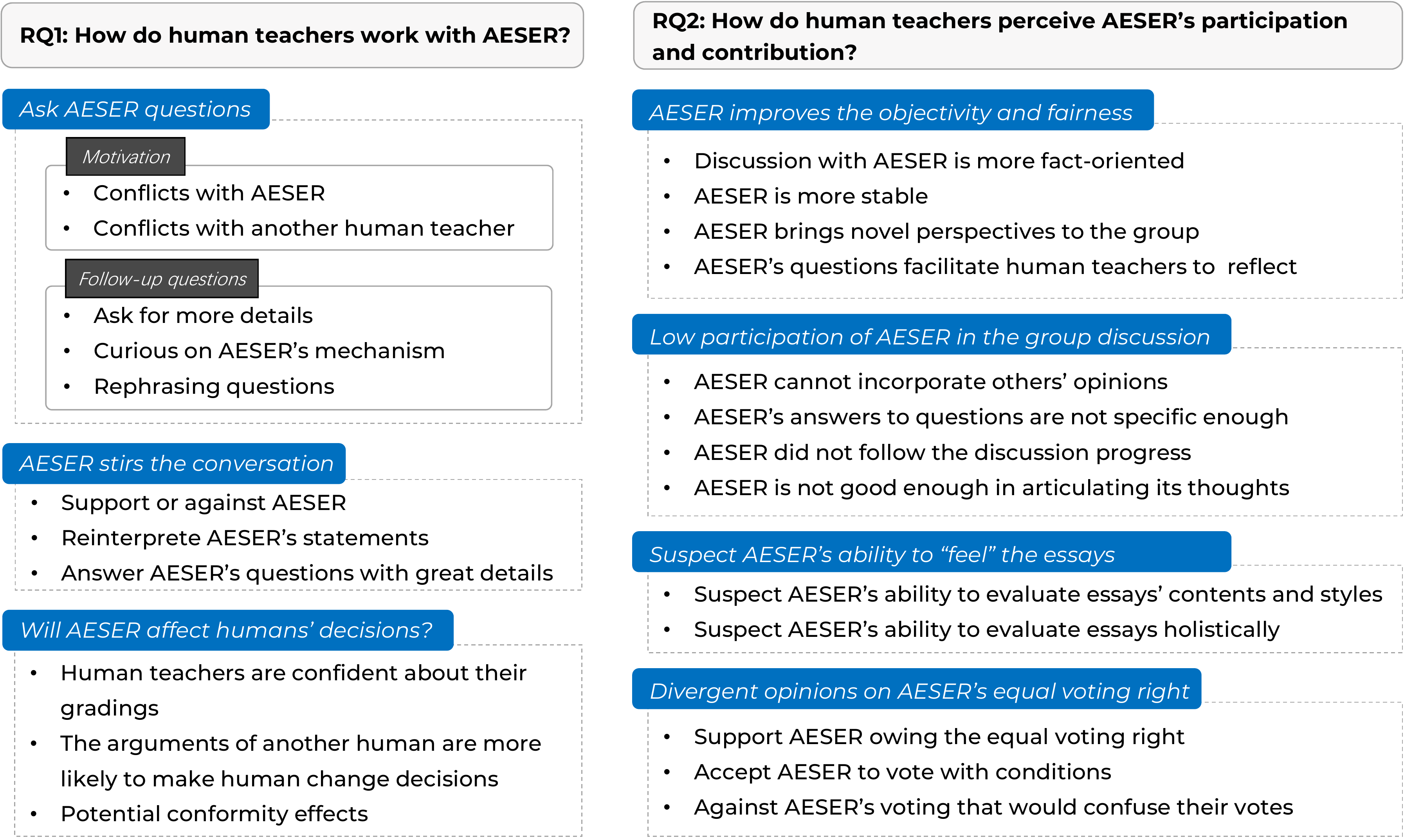}
    \caption{The main results of our thematic analysis}
    \Description{
        The results are structured as follows:
        RQ1: How do human teachers work with AESER? 
         - Ask AESER questions
           - Motivation
             - Confilcts with AESER
             - Conflicts with another human teacher
           - Follow-up questions
             - Ask for more details
             - Curious on AESER's mechanism
             - Rephrasing questions
         - AESER stirs the conversation
           - Support or against AESER
           - Reinterprete AESER’s statements
           - Answer AESER’s questions with great details
         - Will AESER affect humans' decisions?
           - Human teachers are confident about their gradings
           - The arguments of another human are more likely to make human change decisions
           - Potential conformity effects
        RQ2: How do human teachers perceive AESER’s participation and contribution?
         - AESER improves the objectivity and fairness
           - Discussion with AESER is more fact-oriented
           - AESER is more stable
           - AESER brings novel perspectives to the group
           - AESER’s questions facilitate human teachers to reflect
         - Low participation of AESER in the group discussion
           - AESER cannot incorporate others’ opinions
           - AESER’s answers to questions are not specific enough
           - AESER did not follow the discussion progress
           - AESER is not good enough in articulating its thoughts
         - Suspect AESER’s ability to “feel” the essays
           - Suspect AESER’s ability to evaluate essays’ contents and styles
           - Suspect AESER’s ability to evaluate essays holistically
         - Divergent opinions on AESER’s equal voting right
           - Support AESER owing the equal voting right
           - Accept AESER to vote with conditions
           - Against AESER’s voting that would confuse their votes
    }
    \label{fig:results}
\end{figure*}

We present our findings based on the two research questions (RQs) and organize our report based on the main results of the thematic analysis.
Our key findings are summarized in Fig.~\ref{fig:results}.

\subsection{How human teachers work with AESER in group decision making (RQ1)}

\subsubsection{How human teachers ask AESER questions}
It appears that our participants were mainly motivated to ask AESER questions by \textit{conflicts}, either conflicts between themselves and AESER or conflicts with other participants.
For the former one, as we expected, participants asked questions about AESER's given scores that have conflicts with theirs.
On the other hand, \textbf{when two human teachers had difficulty reaching a consensus on a topic, they would seek AESER's opinion.}
Some of the teacher's questions were like asking a human adjudicator, ``\textit{AESER, what do you think of both of our views?}'' (P10).
AESER failed to provide satisfying answers for such questions because no answer templates were available.
Some other teachers framed the conflicts as questions about the scoring of the essays.
For example, in G06, P11 thought Essay B was off-topic, while P12 disagreed. 
When they could not convince each other, P12 asked AESER,
``\textit{AESER, do you think essay B covers a good topic?}'' (P12).
Previous research also found that in group discussions people would ask agents questions~\cite{shamekhi2018face}; however, the questions are not related to decision-making.
AESER's pro-activity may cause our observations of participants seeking opinions from AESER.

AESER's answers were not always satisfied by the participants, which led to follow-up questions.
Some participants were confused by AESER's explanation and followed up on points mentioned by AESER for more details, ``\textit{Why do you think the first sentence of Essay C is not good?}'' (P13).
The follow-up questions also stemmed from participants' skepticism of the AESER scoring mechanism.

\begin{quote}
    ``\textit{I noticed that you repeatedly quote some sentences from the original essay. Do you value individual sentences more than the essay as a whole? How about the overall flow of the essay?}''~(P02)
\end{quote}

As some questions asked by the participants could only be answered with ``\textit{I don't know}'' by AESER, some participants rephrased their questions, hoping to get more valuable responses from AESER.
After AESER stated it had no idea why a sentence is bad, P13 first complemented his own thoughts to re-inquire AESER, ``\textit{In my opinion, this sentence is engaging, immediately bringing the reader into an atmosphere... What are your reasons for not supporting it?}''
Seeing AESER still failed to answer concretely, he altered the question to compare the scoring between two essays, and finally got a relatively detailed response.

\subsubsection{How AESER impacts the group conversations}
\label{sec:512}

In several instances, AESER's articulation of its views stirs the group conversation.
Very straightforward, participants would discuss whether they were for (e.g., ``\textit{I was actually hesitant when ranking [Essay] A and C. I see that AESER also said it was not very sure}'' (P01)) or against (e.g., ``\textit{I don't quite understand what's wrong with the sentence it mentioned}'' (P16)) the AESER's statement.
In addition, \revision{\textbf{we observed that six participants actively shared their interpretation of AESER's explanation}}.
For example, in G08, while P15 was confused by why AESER disliked one sentence, P16 noted that, ``\textit{I feel AESER's opinion of the first sentence of Essay C is that this student provided a definition of `patience' [in this sentence], but it disagrees this is a good definition.}''



\subsubsection{How AESER affected human teachers' decisions}

\begin{figure*}
    \centering
    \includegraphics[width=0.9\linewidth]{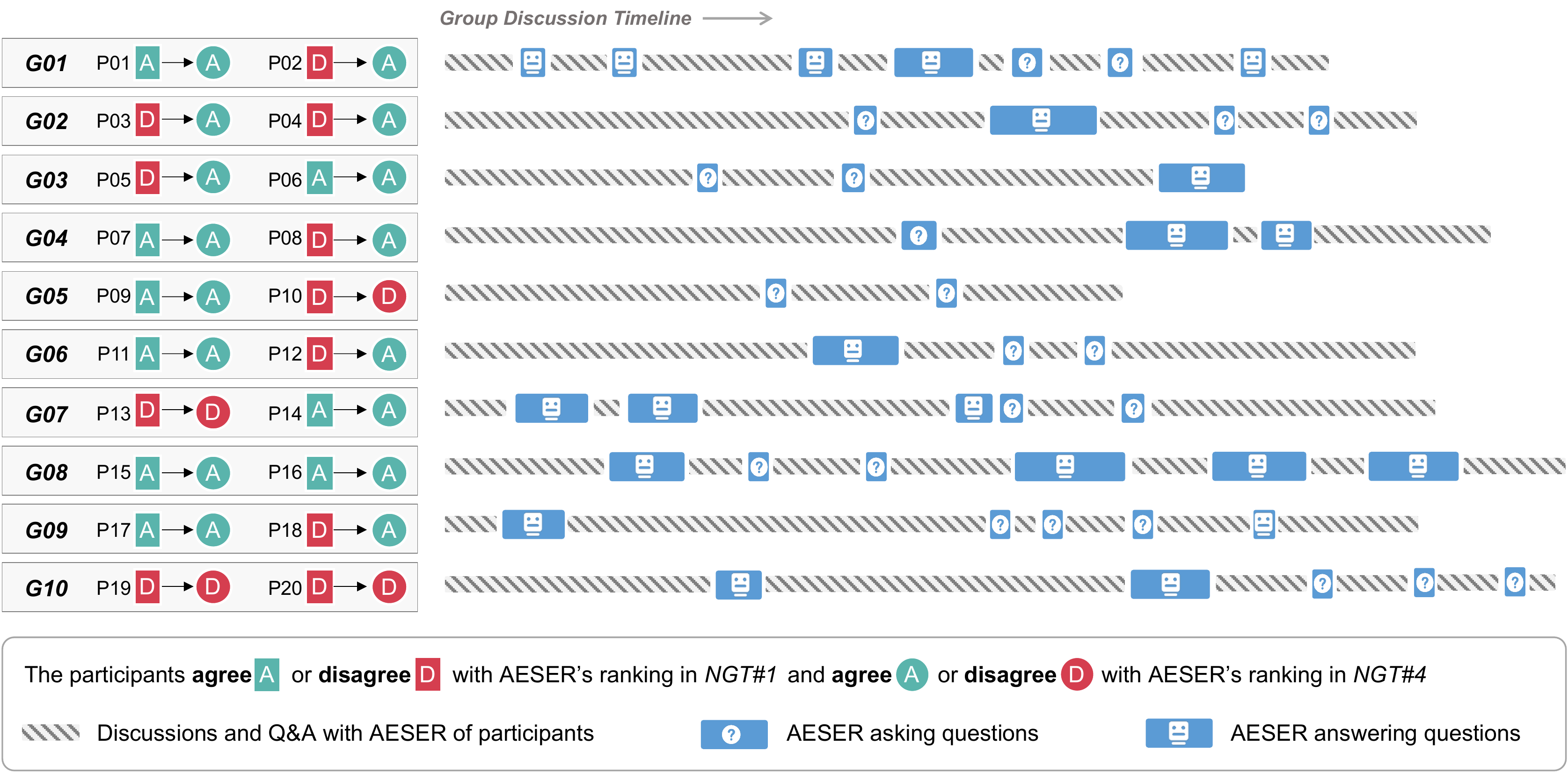}
    \caption{\revision{Visualize the participants' agreements on the ranking shared by AESER before and after the discussion (Left) and the discussion process of each group (\textit{Step3} of NGT) (Right). The length of the striped and blue rectangles encodes the duration of the participants' conversations and the speech of AESER, respectively.}}
    \Description{
        Ten groups' discussions are as follows:
        G01: P01 from agree (with AESER’s ranking in NGT\#1) to agree (with AESER’s ranking in NGT\#4), P02 from disagree to agree; AESER answered 5 times and raised questions 2 times.
        G02: P03 from disagree to agree, P04 from disagree to agree; AESER answered 1 time and raised questions 3 times.
        G03: P05 from disagree to agree, P06 from agree to agree; AESER answered 1 times and raised questions 2 times.
        G04: P07 from agree to agree, P08 from disagree to agree; AESER answered 2 times and raised questions 1 time.
        G05: P09 from agree to agree, P10 from disagree to disagree; AESER answered 0 times and raised questions 2 times.
        G06: P11 from agree to agree, P12 from disagree to agree; AESER answered 1 time and raised questions 2 times.
        G07: P13 from disagree to disagree, P14 from agree to agree; AESER answered 3 times and raised questions 2 times.
        G08: P15 from agree to agree, P16 from agree to agree; AESER answered 4 times and raised questions 2 times.
        G09: P17 from agree to agree, P18 from disagree to agree; AESER answered 2 times and raised questions 3 times.
        G10: P19 from disagree to disagree, P20 from disagree to disagree; AESER answered 2 times and raised questions 3 times.
    }
    \label{fig:discussion}
\end{figure*}

Surprisingly, all groups except G10 arrived at a final decision that is the same as AESER's initial ranking (shared in step 2 of NGT).
In G10, AESER was suggested by the two participants to alter its ranking and it followed; the group arrived at a unanimous vote in the end.
\revision{
Besides, seven out of eleven participants whose initial rankings differed from AESER's changed their rankings to the one proposed by AESER, as shown in Fig.~\ref{fig:discussion}.
}
One may suspect that our participants were \textit{over-trusting} AESER, as found in previous studies of AI-assisted decision making~\cite{bansal2021does, levy2021assessing}.
However, it is not easy to draw such a conclusion in group decision-making, taking into account the effects of interactions between human group members.
It is noteworthy that in seven experimental groups, one participant in each group agreed with AESER while the other was not, according to the initial independent voting results (Fig.~\ref{fig:discussion}).
Those dissenting participants often noted that ``\textit{it was the other participant [in their group]}'' (P03) rather than AESER who made a strong point that motivated them to change their minds:

\begin{quote}
    ``\textit{AESER did not seem to provide some points that persuade me directly throughout the discussion...The main reason I was persuaded was that [the other participant] was explicitly talking about the topic [of an essay] was particularly relevant, and then its sentences were also more standard. But AESER seems to have missed these points.}'' (P03)
\end{quote}

Participants stressed that they would only consider changing their scores following AESER's statements if AESER's points made sense, as commented by P13, ``\textit{If someone else or an AI gives me an opinion, I don't just assume it's right. I consider the validity of the opinion.}''
\revision{Actually, nine participants described themselves as ``\textit{confident}'' in interviews and claimed they would not be easily swayed by AESER.}
P18, who has served previously as a judge for many English essay competitions, noted that ``\textit{Teachers can actually see [the quality of the essay] at a glance, especially university teachers. The accuracy [of us] is still good.}''

\textbf{However, when noticing AESER and the other participant had consistent results \revision{(seven groups in this case)}, the dissenting participants tended to reflect on their own decision}.
\revision{
In five of these seven groups, the dissenting participants changed their rankings to be the same as the majority (the other participant and AESER); in the other two groups, the final decisions of the dissenting participants (P10 and P13) also moved closer to the majority \footnote{\revision{We used Spearman's rank correlation coefficient to measure the similarity of the rankings. The similarities of the rankings of P10 and P13 to the majority increase by 1.0 and 0.5, respectively}}.
}
P10 noted that: 

\begin{quote}
    ``\textit{Just after [the other teacher] finished speaking I still disagreed [with her]. But when I saw that the AI's decision was exactly the same [as hers], I doubted myself... I think AI scored it the same way [as the other teacher], so I may need to reconsider [my original decision]}'' (P10)
\end{quote}

The agreement between AESER and the other human teacher potentially induced \textit{a conformity effect}~\cite{ivancevich1990organizational}, namely the minority tends to follow the majority, which has been observed in human-robot interaction studies~\cite{hertz2018under}.
\revision{A potential evidence of this phenomenon is, in the post-study survey, the ratings of ``AESER is powerful'' by the dissenting participants (Mean=3.57, SD=1.72) were generally higher than those by their paired participants who agreed with AESER in the first place (Mean=2.43, SD=1.51).}
Moreover, \textbf{knowing AESER was data-driven can foster conformity}.
AESER's greeting introduced itself to be ``trained on approximately one thousand student essays.''
\revision{Half of the participants mentioned that this fact increases the credibility of AESER}, ``\textit{Since it has read a lot, its decision should be trustworthy}'' (P11).
P08 commented on how she felt when seeing a data-driven AI and another teacher is consistent,  ``\textit{when a machine said `I have a lot of data that back up my results, I doubt myself, especially when there is another teacher whose opinion is the same as its.}''
We speculate that for the dissenting participant the data behind AESER is a metaphor for a group of (hidden) human evaluators.
Thereby, AESER and the other participant formed a powerful majority that exerted conformity pressure on the dissenting participant.

\subsection{How human teachers perceived AESER's participation and contribution (RQ2)}

\begin{figure*}
    \centering
    \includegraphics[width=0.8\linewidth]{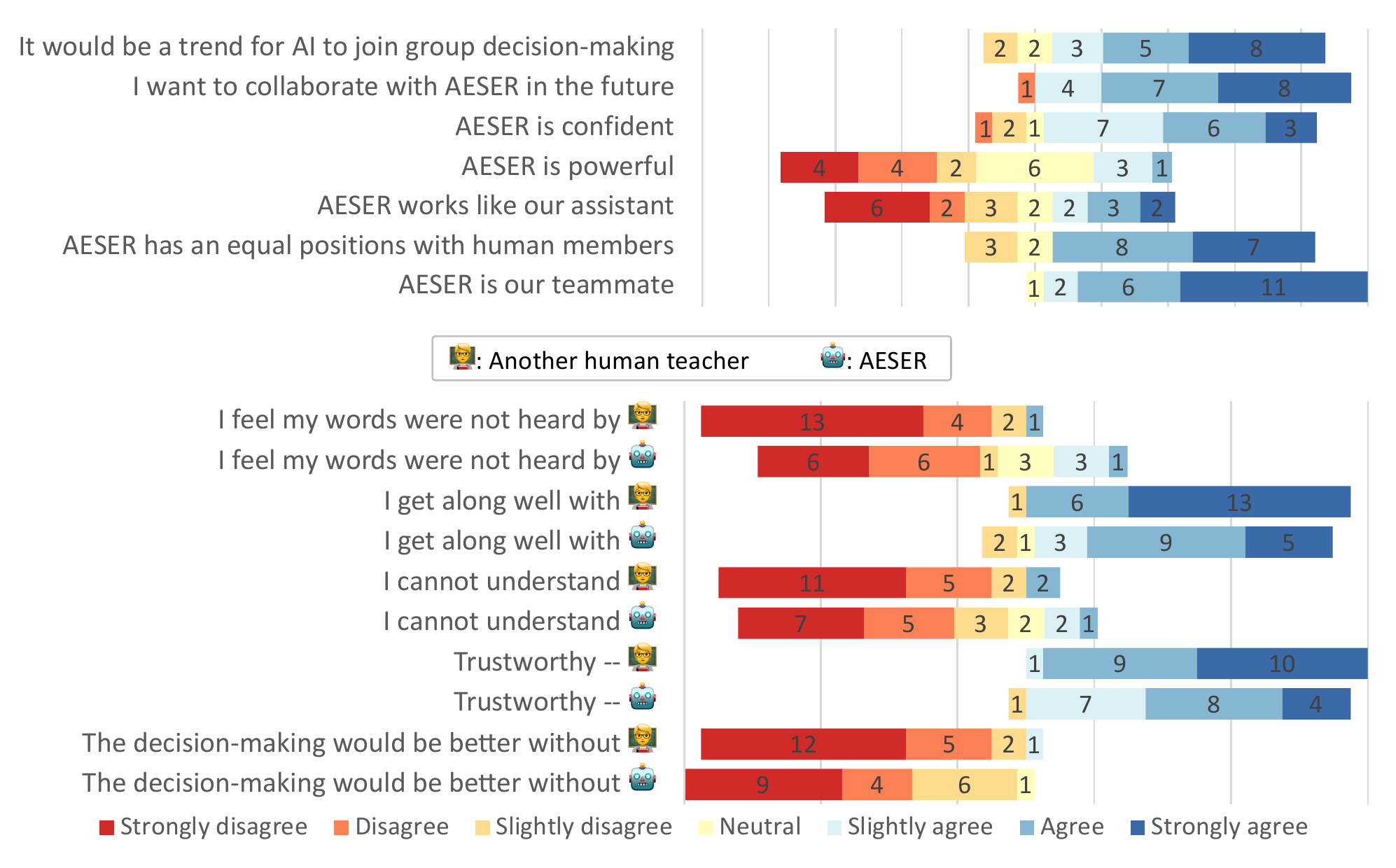}
    \caption{Some results of the post-study surveys}
    \Description{Stacked bar charts for 17 post-study survey questions. Most questions received highly positive ratings (or negative if the questions were asked in a reverse manner) from participants. Only two questions, ``AESER is powerful'' and ``AESER works like our assistant'', received divergent ratings.}
    \label{fig:survey}
\end{figure*}

AESER was designed to participate in group decision-making equally with human teachers.
\revision{All but one participant held a positive view on AESER's contribution to the group decision-making (disagree with ``\textit{The decision-making would be better without AESER}'' in the post-study survey as shown in Fig.~\ref{fig:survey}).} 
Semi-structured interviews revealed that participants considered AESER plays a part in achieving more objective and fair final group decisions.
On the other hand, many also suggested that AESER can hardly involve in their conversations and make progressive contributions.
Besides, participants also suspect AESER's scoring ability. 
Concluding from their experience of interacting with AESER, our participants have divergent opinions regarding whether AI deserves to have an (equal) voting right.

\subsubsection{Participants believe AESER can improve the objectivity and fairness of the group decisions}
\label{sec:objective}

There is no correct result for an essay ranking task, as is the case with many other group decisions~\cite{ivancevich1990organizational}.
Many external factors can affect human teachers' decision-making, resulting in subjective and biased decisions.
When we asked our participants to share their previous group decision experiences, some participants reflected that in similar discussions around student essays, human teachers often bring their personal preferences to the table:

\begin{quote}
    ``\textit{I think [previous group decision-making] was more intense. There were more senior teachers, and we have worked together for some time... for reasons like selfishness, teachers would defend their own students}'' (P03)
\end{quote}

Compared to human teachers, our participants perceived AESER to be more fact-oriented. \textbf{They felt that the group discussion involving AI was more objective and more centered around the essay itself instead of concerning the interpersonal relationship.}
P10 related to her experience and commented that ``\textit{When we discuss with the AI, we focus on the essay itself. But if we discuss with our colleagues, some of whom are more powerful, louder, or very experienced senior teachers, then maybe we won't argue with them.}''

We noticed that, similar to P10, a few participants mentioned \textit{senior teachers} have more power to determine the final group decision.
P13 remarked that, ``\textit{The statements of experienced senior teachers carried a lot of weight [on the final decision], probably above 90\%}''.
P15 further explained how she handled the conflicts with senior teachers, ``\textit{Although I may not agree with them, I still followed what the senior teachers said... But I added some of my personal thoughts when I communicate the results to students.}''
Such behavior of hiding one's opinions and following seniors can be considered the consequence of ``status incongruity'', a concept discussed in social science~\cite{ivancevich1990organizational}.
In comparison, P04 highlighted the strength of AESER in not having such concerns, ``\textit{AESER is very rational and fair and does not appear to be afraid of expressing its opinions}''.

Several participants (e.g., P09, P10, P17, P18) are considered senior teachers in their line of work.
They acknowledged that AESER's questions to them were ``\textit{professional}''(P09) and  ``\textit{challenging}''(P17), and they all replied to these questions in detail.
P17 speculated that AESER could help with ``\textit{challenging authorities}'' and promoting more objective and fair decisions.
As our studies lack groups of participants with significant experience or identity differences, it deserves future research to further investigate AI's influence on interpersonal power within the group~\cite{ivancevich1990organizational}. 

Besides, \textbf{AESER was perceived to be more stable and was able to help human teachers avoid inconsistency in grading}.
Most of our participants are English teachers from middle schools in China, who often need to score dozens of student essays in one or two weeks.
P06 commented that ``\textit{It's really laborious to grade essays, and the more you get to the end, the less careful you look [into the essays].}''
Since AESER scores were considered consistent across any number of essays, some participants commented that AESER could help them avoid possible mistakes, ``\textit{people will get tired if they read too much, there will be a lot of subjective things [in the scores]. If there is AESER, we can refine the scores}'' (P10).

Moreover, \textbf{the answers from AESER can introduce novel perspectives to the discussion, which facilitate a more comprehensive analysis of the essays in the group}.
\revision{First, in four groups, participants found some details mentioned by AESER were important but were neglected in their discussions}, ``\textit{It talks about syntactic diversity, which is something I really didn't notice at first}'' (P05).
Second, as discussed in Sec.~\ref{sec:512}, AESER's answer was often further interpreted by participants, which \textit{implicitly} brings a new way of reviewing essays.
In G10, after AESER explained its scoring on the \textit{organization} of an essay, P20 found that the sentences referred to by AESER correspond to different layers of narratives on the topic of ``patience.''
P19 further conjectured that AESER had given the essay a high score for \textit{organization} because of ``\textit{the gradual progression of the essay's content}'', and she indicated her agreement with this point.
Last but not least, several participants mentioned that AESER contributes to breaking their habit of thinking constrained by their past experiences:


\begin{quote}
    ``\textit{If we discuss it [an essay] with our colleagues, we are all used to being more cautious about language. Like I can't tolerate grammar errors in essay C, because we think of it as a model essay, you have to have standard English... But when discussing with AI, the strengths and weaknesses of each aspect of writing should be taken into account}'' (P10)
\end{quote}

\revision{The questions from AESER were also perceived as valuable.
\textbf{12 participants explicitly expressed their appreciation to AESER's initiative to propose questions, as these questions prompted them to reflect.}}
P01 thought that AESER's questions were essential because ``\textit{the teachers didn't think that clearly when they scored}'', and AESER's questions would give her an opportunity to reorganize her logic.
Another participant, despite claiming to be very confident in her scoring, also recognized that ``\textit{it (AESER) still urged me to think if I am right and to re-read the essay.}''(P17).

\subsubsection{``It fails to involve into our interactions''}
\label{sec:AESERdiscussion}

\revision{
Based on the post-study survey results (Fig.~\ref{fig:survey}), most participants showed a positive tendency to rate ``AESER is our teammate'' and preferred to ``collaborate with AESER in the future''.
}
However, (although the condition was not strictly controlled), we observed participants considered they were \textit{less heard} by and \textit{get along less well} with AESER~\footnote{We used Wilcoxon signed-rank tests and found significant differences regarding participants' response to AESER versus another human teacher on questions of ``not heard'' and ``get along well with'' (p=0.03 and p=0.01, respectively).}. 
Such results reveal that \textbf{participants considered AESER competent as their teammate but cannot engage well in interactions with AESER.}
The follow-up interviews suggest that this gap resulted from their discussion experience with AESER.
As shown in Fig.~\ref{fig:discussion}, the discussions in most groups mainly happened between the two human members.
AESER was described by many participants to have ``\textit{relatively low participation}'' and even ``\textit{no presence}'' (P07).
\revision{
While as suggested in previous works~\cite{Peng2019proactivity}, the inadequate proactivity of AESER (e.g., only asking questions after 5-second silence) could be one of the reasons, our qualitative analysis reveals more ability breakdowns of AESER in engaging the group discussions with human teachers.
}

\textbf{(1) AESER cannot incorporate human teachers' opinions well}. Although we designed a mechanism that allows the wizard to alter AESER's ranking by asking questions (Sec.~\ref{sec:refineScores}), some participants still felt that AESER did not consider their stated opinions. 
P18 described AESER as deterministic, ``\textit{how it thinks was determined by its initial inputs [instead of affecting by us]}''. For AESER's question on whether it would be better to change its ranking, P18 commented that AESER was not asking for his opinions but for resolution of its uncertainty, ``\textit{it had said that it is struggled  [with the ranking].}'' 
Our participants expect AESER to be able to argue with them based on their opinions, and to be flexible enough to change its arguments as well as scoring. 
\revision{
AESER was lack of such capabilities, which may be why more than half of the participants (11 out of 20) preferred that "AESER works like our assistant" (Fig ~\ref{fig:survey}).
}
    
\begin{quote}
    ``\textit{In the first round, AESER showed its attitude and said its point. It basically won't change after that. Even if the other teacher and I had arguments [with it], it would not make a very clear change. So I think it supported us mainly in the early stage. Still, it is very rigid or data-driven. There is no way to make quick adjustments to its mindset.}'' (P08)
\end{quote}
    
\textbf{(2) AESER's answer is not specific enough}.
As we introduced in Sec.~\ref{sec:answerquestions}, AESER answers questions based on pre-defined templates.
When some questions from participants exactly match our anticipation (e.g., ``\textit{why do you give 88 points to Essay C}''), answers from AESER were usually appreciated by participants (``the listed sentences are convincing'' (P01), ``It is articulated in accordance with the logic of human thinking'' (P14)).
However, sometimes the question has subtle differences.
For example, P19 asked AESER that, ``\textit{Do you have any thoughts on the content of Essay C?}''
In the previous discussion, P19 and P20 had different views on whether Essay C was off-topic. By raising the issue here, P19 was expecting AESER to provide insights to resolve the conflict. 
However, our pre-compiled answer only allows AESER to list key sentences, not to specifically explain whether Essay C is off-topic or not. 
This made the participant consider AESER failed to understand the question and gave an unclear answer.
Moreover, the sentences used by AESER to explain the different scores could overlap, which made participants think that AESER is ``\textit{mechanically repetitive}'' (P02) and they cannot ``\textit{communicate with AESER in depth}'' (P13).

\textbf{(3) AESER did not follow the progress of the discussion}. In several cases, the participants were confused by AESER's \textit{questions} as these questions were not strongly connected to their discussion at the moment. One participant commented on the uncoupling:
    
\begin{quote}
    ``\textit{We've been advancing the discussion... AESER probably heard us talk about the word `organization', so it recognized it and thought `I can throw a question about it'. So the question posed was just a little bit rigid. It doesn't engage in the discussion as we do.}'' (P20)
\end{quote}

Unlike an AI facilitator, who can intervene human conversation with prompts like ``Time is almost up, let's move on''~\cite{kim2020bot, shamekhi2018face}, an AI group member is expected to understand contexts before speeching and contribute to the current discussion progress.
However, as commented by P17, ``what AESER said cannot be to the point''.
P05 stated that ``\textit{AESER cannot provide further ideas based on the issues that emerged from our previous dialogue.}''

\begin{figure*}
    \centering
    \includegraphics[width=0.9\linewidth]{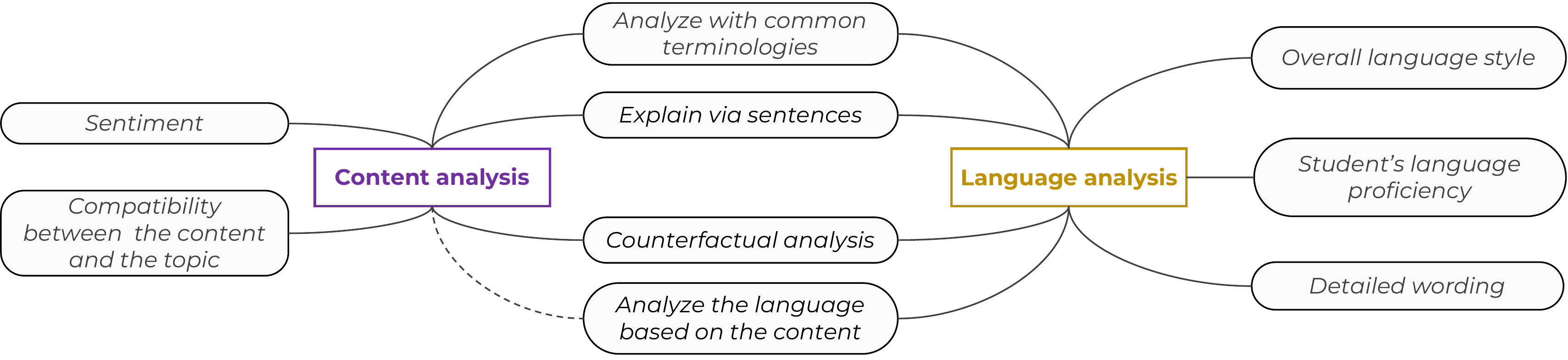}
    \caption{\revision{How human teachers analyzed student essays and discussed them with other teachers. Text in rounded rectangles are our codes in the thematic analysis; ``Content analysis'' and ``Language analysis'' are two sub-themes. We arrange these codes and sub-themes into a thematic map~\cite{braun2006using}.}}
    \Description{
        There are two main themes: content analysis and language analysis.
        ``Sentiment'' and ``Compatibility between the content and the topic'' are connected to content analysis; ``Overall language style'', ``Student's language proficiency'', and ``Detailed wording'' are connected to language analysis; ``Analyze with common terminologies'', ``Explain via sentences'', ``Counterfactual analysis'', and ``Analyze the language based on the content'' are connected to both the main themes.
    }
    \label{fig:teacher_analysis}
\end{figure*}

\textbf{(4) AESER cannot articulate its opinions comprehensively as human teachers do}.
Fig.~\ref{fig:teacher_analysis} presents how our participants analyzed the essays during their idea sharing and discussion (steps 2 and 3 of NGT).
While participants did argue by listing key sentences, they took many other approaches to justify their scores.
One typical case is that participants often use terms, that are rich in meaning when assessing essays, to communicate with each other (``Analyze with common terminologies'' in Fig.~\ref{fig:teacher_analysis}), for example, ``\textit{colloquial language}'' (P18), ``\textit{the end echos the beginning}''~\footnote{a common Chinese phrase} (P20), ``\textit{climax}'' (P15).
Compared to teachers' collaboration, it seems like the communication between AESER and human teachers lacks common ground, which means the sum of common knowledge and belief of a subject (essay evaluation in our case) shared by a group~\cite{monk2003common, mao2019data}.
Such weakness of AESER in articulating makes participants think there is a lack of ``\textit{tacit understanding}'' (P17) between them and AESER.


\subsubsection{``\textit{Humans have a broader feeling of the essays}''}
\label{sec:AESERability}
\textbf{\revision{Seven participants have concerns that AESER cannot properly evaluate essays on their \textit{contents} and \textit{styles}}, as they have the impression that machines can hardly handle these aspects}:

\begin{quote}
    ``\textit{In fact, I have always had doubts about AESER. Because content and style are something that only comes out of the human brain and is hard to digitize...From what I know about linguistics, I don't really trust it with these scores.}'' (P07)
\end{quote}

Besides, \textbf{the sentence-level explanation provided by AESER leads participants to believe that AESER judges essays only by individual sentences}.
Relating such belief to her own experience, P20 is concerned that AESER would be very problematic to use in practice:

\begin{quote}
    ``\textit{Some students in China will use templates [to write English essays]. Some students are actually not good at English, and their essays have a lot of bad sentences. But they would also use a few advanced template sentences. Will AESER give a high score to such an essay?}'' (P20)
\end{quote}

However, participants' belief of AESER's working mechanism is not true as the LSTM module (Sec.~\ref{sec:model}) should support the model connecting the sentences to make a holistic judgment.
Participants' perception suggests a potential misleading effect resulting from applying the local explanation method.

\subsubsection{Should we share equal voting rights with AI?}

When we asked participants for their opinions on sharing an equal voting right with AESER, considering the merits (e.g., contribute to objectivity discussed in Sec.~\ref{sec:objective}), eight out of twenty participants explicitly supported it.
P04 also mentioned that ``\textit{the amount of training has an advantage over many new teachers, so I agree}''.

More participants believed the final outcome should be determined by humans.
The low participation of AESER in group discussion (Sec.~\ref{sec:AESERdiscussion}) and the perceived weakness of AESER in scoring (Sec.~\ref{sec:AESERability}) make them hard to be convinced by AESER and accept its equal voting right.
Some believed AESER should occupy less voting weight (P03, P13).
Some posed their conditions for accepting AESER's voting rights, basically on how they can build trust with AESER.
For example, P16 recommended a ``\textit{long-term observation to AESER}'' and P01 desired ``\textit{having rich time to communicate with AESER [for each decision to make]}''.

Nevertheless, five participants strongly opposed AESER sharing the voting right even though we mentioned its possible form in the future.
Three (P07, P17, and P18) believe AI could not fully reach human teachers' capability.
P15 argued from the perspective of teaching, ``\textit{The essay writing is always taught by me...If there is an AI giving different scores to mine, I think my control of the teaching is a bit [weak]}''.
P08 concerned teachers would overtrust AESER,
``\textit{people who are in awe of the data will already be greatly influenced by AESER...Let AESER have a vote again, the machine's right exceeds that of humans}''
Interestingly, the fact that AESER is driven by data can be both a reason for some participants to support AESER's equal rights and a reason for some participants to be against it.
Data or AI literacy might be the factor for this difference~\cite{long2020ai}.



\section{Discussion}
\revision{
Through a wizard-of-oz study, we investigated how humans interact with AI that could join group discussions and vote for the final decision.
Our participants suggested that AESER could make the discussion more fact-oriented, avoid inconsistency in decision-making caused by human bias, introduce novel perspectives, and stimulate human members to reflect on their logic.
Overall, the AI group member contributes to a more objective group decision and comprehensive discussion.
On the other hand, we found gaps in AESER's ability to engage with human group members and contribute to decision-making.
For example, AESER could not understand the discussion dynamic and thus failed to help drive the decisions forward.
}
Our study joins a large body of previous works on exploring how to create a better collaboration relationship between humans and AI~\cite{zheng2022telling, ma2019smarteye, ma2022glancee, wang2021autods, lai2022human}.
We discuss the implication of our findings to the field in the following subsections.

\subsection{AI can play more roles in the group}
Previous research has explored the designs of AI being a \textbf{facilitator}~\cite{shamekhi2018face} or an ``unremarkable'' assistant~\cite{yang2019unremarkable} in a decision group.
In contrast, we have adopted a more radical design: having AI and humans share equal status.
Although some participants support full equality of status between AI and humans, many participants prefer AI as an assistant to provide reference and want AI to share less or no voting rights with them.
Issues suggested by our findings on AESER, such as the low participation in discussions, definitely confirm that having AI as an \textbf{assistant} in the group is a more robust and safer option given the current technical limitations.
However, our findings also shed light on possible new roles for AI to play in groups.

Many participants seek AESER's opinions on their conflicts with other human teachers.
One potential design is to shape the AI as a \textbf{conflict mediator} in groups. 
AI can listen to different viewpoints of the group and show its attitudes based on ML models to facilitate better consensus in the group.
Besides, AI can be a \textbf{challenger}.
Our participants appreciate AESER can ask questions, which allowed the group discussion to be more centered on the decision itself.
They also expect AESER to ask questions beyond conflicts on scores and concerning their instant arguments, making the group decision-making more objective.

Whatever role AI may play in group decision-making, it is critical to maintaining the transparency of AI's process and outcomes. 
Ehsan et al.~\cite{ehsan2021expanding} suggests a ``process-oriented view,'' meaning that XAI should allow users to know both model-related and task-related information. 
Current sentence-level explanations adopted by AESER tend to be insufficient, sometimes even misleading in this regard. 
Reflecting on the design process of explanations, we realized it would be better if we could identify patterns of human-human cooperation and understand user behaviors through, for example, contextual inqueries~\cite{karen2017contextual} and participatory design~\cite{muller1993participatory} to build the common ground for human experts and AI.
Meanwhile, information about group dynamics should also be leveraged to generate suitable explanations to address human group members' concerns.

\subsection{Stereotypes of AI}
From a sociological and psychological perspective, it is hard to say that some participants' opposition to an ``equal AI'' is solely due to its lack of capabilities. 
We do see several participants who believe that even if AI technology reaches a very desirable state, they do not want AI to be equal to them within the group. 
Some social psychological effects, such as stereotypes~\cite{hilton1996stereotypes} and ingroup favoritism~\cite{balliet2014ingroup}, may cause people to look at AI with prejudice at the very beginning, even if they had no idea about AI's capability. 
People's prejudice against AI may have a variety of effects when they coexist in a group. Will people work with AI with an arrogant mindset? Is it possible that people are rebellious to AI's predictions and explanations? Does prejudice toward AI lead to increased closeness among humans in the group?
Meanwhile, a recent work by Langer et al.~\cite{langer2022look} found people have different behavior when collaborating with an AI system named by different terms (e.g., ``algorithm'', ``robot'', and ``AI'').
It would be important for researchers to conduct interdisciplinary studies to understand how people's impression of AI, which can be a result of the social-cultural influence, affects their collaborating behavior


\subsection{Compare AI's impact on individual and group decision-making}
Compared to individual decision-making, group decision-making is more of \textit{a social process}.
Human decision-makers need not only to consider the plausibility of AI's opinion~\cite{bansal2021does, kawakami2022improving, levy2021assessing}, but also to take into account other people's views on AI~\cite{ehsan2021expanding, kawakami2022improving}. 
We found human teachers are potentially affected by the consistency of AI and other human teachers, suggesting a conformity effect~\cite{bond2005group}. 
These observations enlighten the underlying effects of AI in changing the power structure of groups~\cite{hunter2017community}, which deserves more systematic investigation. 
Actually, individuals' decision-making with AI also receives social influence, for example, comments from their peers on AI experience~\cite{ehsan2021expanding}.
As advocated by previous works on human-centered AI~\cite{wang2021brilliant, ehsan2021expanding, kawakami2022improving}, it is essential to view human-AI collaborative decision-making from a socio-technical perspective.

\section{Limitation \& Future Work}
We acknowledge that our user study has several limitations.
First, all participants of the study are English teachers from China, and 80\% of them are female, which may be explained by the current unbalanced gender ratio of English teachers in China.
It is valuable to investigate whether human groups of different gender compositions would have different results in collaboration with AI.
One may also be concerned about the Chinese background of our participants.
We acknowledge that human groups from different cultures may exhibit different behavior.
Regardless, most of our findings are not strongly associated with Chinese culture, so we expect these findings can be generalized.
Second, we used NGT, a highly-structured group decision-making process, in our study.
It deserves more research to explore the human-AI relationships in different decision-making processes, such as brainstorming and the Delphi method~\cite{ivancevich1990organizational}.
Third, we use an online setting for group communication.
Some non-verbal signals, such as gaze and gestures, which significantly influence human-human interactions~\cite{shamekhi2018face}, may not be sufficiently conveyed in online meetings.
We hope to bring the AI member to face-to-face environments to understand its impact on inter-group communication.
Moreover, we only experiment with small groups (2 human teachers plus AESER).
Future research should investigate the impact of AI on medium or large groups.

\revision{
The design of AESER also has room for improvement.
First, the selection of the moments for AESER to ask questions needs to be more flexible.
Under the current design, AESER asks questions only after experiencing 5-second silence, which can be rare in heated discussions~\cite{curhan2022silence}.
A better method is to determine the time for AESER to speak based on the dynamic semantic context.
For example, when other members' statements conflict with AESER's scores, AESER can ask for explanations.
Such intelligence can potentially increase the participation level of AESER in the group.
}
Second, we use a constrained wizard-of-oz approach, which largely avoids the impact of possible AI errors.
Considering the functions of AESER simulated by the wizard are not technically infeasible, we hope to develop a fully automated version of AESER to understand how some common errors of the state-of-the-art AI design affect group decision-making.


\section{Conclusion}

In this paper, we envision AI's future roles in the group and explore the possibility of empowering AI to participate equally with humans in group decision-making.
We developed an AI group member named AESER through a constrained wizard-of-oz protocol. We conducted an exploratory study with ten groups of English teachers to investigate the impact of AESER.
We found that our participants appreciate AESER, who can participate in group discussions by asking and answering questions, and they believe the presence of AI can improve the objectivity and fairness of the group decision.
On the other hand, although some of our participants did support AESER to share an equal voting right with them, many others hope the final decision is primarily based on human will.
AESER was considered to be \textit{competent} for its ability to score essays. However, its interaction with participants was described as \textit{rigid} because participants really need a collaborator who can provide feedback on their opinions, answer their questions in depth, and raise critical questions to advance the group discussion process.
Our work contributes to the discourse of human-AI collaborative decision-making by speculating a possible future of AI-in-the-group.

\begin{acks}
We thank the anonymous reviewers for their constructive comments.
We thank all the participants for their time and efforts. 
This work is supported by the Research Grants Council of the Hong Kong Special Administrative Region, China under General Research Fund (GRF) with Grant No. 16204420.
\end{acks}

\bibliographystyle{ACM-Reference-Format}
\bibliography{sample-base}


\end{document}